\documentclass[onecolumn,aps,prd,preprintnumbers,showpacs,superscriptaddress,nofootinbib,amsmath,amssymb,floats,floatfix,showkeys,notitlepage,longbibliography]{revtex4-1}

\usepackage{orcidlink}
\usepackage{lipsum}
\usepackage{graphicx}
\usepackage{subfigure}
\usepackage[utf8]{inputenc}
\usepackage{palatino}
\usepackage{changes}
\usepackage{hyperref}
\hypersetup{colorlinks=true,linkcolor=blue,urlcolor=blue,citecolor=blue}
\usepackage[toc,page]{appendix}
\usepackage[normalem]{ulem}
\usepackage{adjustbox}
\usepackage{latexsym}
\usepackage{amsmath}
\usepackage{amssymb}
\usepackage{amsfonts}
\usepackage{dcolumn}
\usepackage{bm}
\usepackage{tikz}
\usepackage{bigints}
\usepackage{array,tabularx,multirow,booktabs}
\usepackage[tracking=true]{microtype}
\SetTracking{}{500}
\SetTracking{encoding={*}, shape=sc}{40}
\UseRawInputEncoding 
\allowdisplaybreaks

\newcommand\fc{\frac}
\newcommand\lt{\left}
\newcommand\rt{\right}

\begin{document} \sloppy

\title{Charged black holes in Kalb-Ramond gravity:
Weak Deflection Angle, Shadow cast, Quasinormal Modes and Neutrino annihilation}

\author{Reggie C. Pantig
\orcidlink{0000-0002-3101-8591}
}
\email{rcpantig@mapua.edu.ph}
\affiliation{Physics Department, Map\'ua University, 658 Muralla St., Intramuros, Manila 1002, Philippines}

\author{Ali \"Ovg\"un
\orcidlink{0000-0002-9889-342X}
}
\email{ali.ovgun@emu.edu.tr}

\affiliation{Physics Department, Eastern Mediterranean
University, Famagusta, 99628 North Cyprus, via Mersin 10, Turkey}

\author{\'Angel Rinc\'on
\orcidlink{0000-0001-8069-9162}
}
\email{angel.rincon@ua.es}
\affiliation{Departamento de Física Aplicada, Universidad de Alicante, Campus de San Vicente del Raspeig, E-03690 Alicante, Spain}
\affiliation{Research Centre for Theoretical Physics and Astrophysics, Institute of Physics, Silesian University in Opava, Bezrucovo nám. 13, 74601 Opava, Czech Republic}

\date{\today}
\begin{abstract}
In this paper, we investigate the phenomenology of electrically charged black holes in a Lorentz-violating gravitational framework mediated by a background Kalb-Ramond (KR) antisymmetric tensor field. Employing the Gauss-Bonnet theorem in a non-asymptotically flat geometry, we derive analytic expressions for the weak deflection angle of light and massive particles, revealing persistent corrections due to the Lorentz-violating parameter $ \ell $. 
Scalar and Dirac perturbations are also studied using both the WKB approximation and the Poschl-Teller approximation approach to verify the stability of the solution against these types of perturbations.
Shadow analysis further uncovers a nontrivial deformation of the photon sphere and critical impact parameter, with KR-induced effects modifying the charge contribution in a manner incompatible with standard Einstein-Maxwell theory. Constraints derived from Event Horizon Telescope data for Sgr A* and M87* validate the model and provide stringent bounds on $ \ell $, establishing the KR framework as an observationally testable extension of General Relativity.


\end{abstract}



\maketitle

\section{Introduction}
The exploration of gravitational theories that extend or deviate from General Relativity (GR) has gained notable momentum, especially in regimes where quantum and relativistic effects intersect. One particularly compelling avenue involves the incorporation of spontaneous Lorentz symmetry breaking (LSB) through the dynamics of the Kalb-Ramond (KR) field, a rank-2 antisymmetric tensor field originally introduced in the context of string theory by Kalb and Ramond \cite{Kalb:1974yc} to generalize gauge invariance beyond vector fields. This field naturally emerges in the low-energy effective actions of heterotic string theories and plays a pivotal role in several gravitational extensions that postulate a background tensor field with a fixed vacuum expectation value, thereby violating local Lorentz invariance without explicitly breaking diffeomorphism symmetry.

The theoretical significance of LSB induced by a KR background lies in its ability to probe the foundational symmetry structure of spacetime. When coupled non-minimally to curvature invariants or gauge fields, the KR field gives rise to rich gravitational dynamics, including modified black hole solutions, altered causal structures, and testable deviations in light propagation. The seminal work of Kosteleck\'y and Samuel \cite{Kostelecky:1988zi}, and its subsequent formalization in the Standard-Model Extension (SME) framework, paved the way for such investigations, providing a robust theoretical infrastructure for spontaneous Lorentz violation in both the gravitational and gauge sectors.

Black holes in these Lorentz-violating (LV) scenarios have drawn particular attention, not only due to their theoretical richness but also because of their capacity to manifest observable imprints. Notably, studies have extended to KR-deformed metrics where charged or rotating black hole solutions differ from the classical Reissner-Nordstr\"om or Kerr geometries in both qualitative and quantitative ways. These deformations offer potential insights into quantum gravity, gauge field interactions, and the structure of the vacuum.

The study of black hole shadows has emerged as a vital avenue in testing modifications to General Relativity (GR), particularly in theories incorporating Lorentz invariance violation (LIV). Observational efforts by the Event Horizon Telescope (EHT) have invigorated this line of inquiry, prompting a surge of theoretical developments that explore how LIV alters photon trajectories and, thereby, the morphology of black hole shadows. In the context of Kalb-Ramond gravity, an antisymmetric tensor field introduces corrections to the Kerr spacetime, yielding distinguishable shadow deformations and reductions in apparent size with increasing LIV parameter values \cite{Liu:2024lve}. A similar pattern emerges in charged Kalb-Ramond black holes, where the shadow and innermost stable circular orbit (ISCO) exhibit sensitivity to the LIV parameter $ \ell $, decreasing as $ \ell $ increases \cite{Duan:2023gng}. Moreover, in rotating Kalb-Ramond geometries, increasing the field strength induces significant distortion and diminishment of the shadow radius, effects that are consistent with EHT measurements of M87* and Sgr A* \cite{Kumar:2020hgm}. On the bumblebee gravity side—a vector-tensor extension that facilitates spontaneous Lorentz symmetry breaking—analogous modifications arise. The Lorentz-violating parameter $ \ell $ influences both weak and strong field deflection angles and significantly alters the shadow geometry in Reissner-Nordström and Kerr-like spacetimes, often producing enhanced shadow radii and greater oblateness, particularly for negative values of $ \ell $ \cite{Lambiase:2024uzy,Islam:2024sph}. Observationally grounded studies have now proposed precise constraints on these LIV parameters, with upper bounds on $ \ell $ derived through EHT-based shadow analysis and Solar System tests ranging from $ 10^{-4} $ to $ 10^{-9} $, depending on the gravitational model and metric considered \cite{Fathi:2025byw,Alimova:2024bjd}. Investigations such as those by Pantig and \"Ovg\"un \cite{Pantig:2025paj} analyzed the modifications of photon spheres and shadow morphology in asymptotically AdS Kalb-Ramond black holes, revealing constraints on the LV parameters through high-resolution imaging like that of the Event Horizon Telescope. Then, there is also the deflection of light around KR black holes departs from classical predictions. Notable contributions include works by Filho et al. \cite{Filho:2024tgy} and Atamurotov et al. \cite{Atamurotov:2022wsr}, which quantified the lensing observables and time delay effects arising from antisymmetric tensor backgrounds. The thermodynamic response of black holes in KR gravity also deviates from GR predictions. Filho et al. \cite{AraujoFilho:2024rcr} investigated evaporation rates, deflection angles, and the interplay between QNMs and arguing that these phenomena could collectively signal underlying Lorentz-violating dynamics. The critical structure of photon orbits and their thermodynamic analogs have been extended to KR geometries. Liu et al. \cite{Liu:2024oas} revealed topological signatures encoded in the phase diagrams of black hole shadows and thermal entropy. Recent studies have extensively investigated gravitational lensing, thermodynamics, and optical properties of black holes influenced by the Kalb-Ramond field in Lorentz-violating contexts~\cite{Mangut:2025gie, Ovgun:2025ctx, Tan:2025pya, Cordeiro:2025cfo, Sucu:2025lqa, Belchior:2025xam, Jumaniyozov:2025wcs, Alimova:2024bjd, Ma:2024ets, al-Badawi:2024pdx, Zahid:2024ohn, Hosseinifar:2024wwe, AraujoFilho:2024rcr, Du:2024uhd}.

Beyond shadow observations, comparative and theoretical studies reveal distinctive phenomenological fingerprints of LIV in both Kalb-Ramond and bumblebee frameworks. The thermodynamic behavior of AdS black holes, analyzed through free energy landscapes and Ruppeiner geometry, displays model-dependent phase transitions, including overlapping thermodynamic regimes and multiple critical points induced by the nature and magnitude of Lorentz-violating fields \cite{Masood:2024oej}. A broader comparative analysis highlights how observable quantities such as photon radii, Hawking temperature, and quasinormal modes diverge between the Kalb-Ramond and bumblebee scenarios, with Kalb-Ramond black holes typically exhibiting enhanced Hawking radiation and greater electromagnetic wave transmission compared to their bumblebee or Schwarzschild counterparts \cite{Jha:2024xtr}. These distinctions are not merely theoretical but find relevance in gravitational wave physics as well, Kalb-Ramond black holes produce faster-decaying ringdown signals in scalar and electromagnetic perturbations, thereby offering a potential discriminant in future LIGO/Virgo detections. From a geometrical perspective, the introduction of conical defects in Schwarzschild-like bumblebee spacetimes leads to modified lensing formulas, where angular separations and flux ratios deviate exponentially with LIV parameters \cite{Izmailov:2022jon}. Such geometrical features are further compounded when global monopoles are considered, introducing topological contributions to the metric that enhance both shadow size and deflection angles \cite{Gullu:2020qzu}. Furthermore, analyses of Kerr-Sen-like and rotating bumblebee black holes show that LIV-induced deformations persist even in plasma environments, reinforcing the robustness of these effects across different astrophysical conditions \cite{Jha:2021eww}. Collectively, these findings suggest that a synthesis of thermodynamic, dynamical, and observational data can effectively isolate the signatures of Lorentz-violating extensions of gravity, providing a comprehensive framework for their empirical validation.

Quasinormal modes (QNMs hereafter) of black holes are intrinsic oscillations that arise when a black hole is perturbed, which eventually decaying due to gravitational wave emission. These modes are solutions to the linearized perturbation equations in a black hole spacetime, characterized by complex frequencies $\omega \equiv \omega_R + i \omega_I$, where the real part $\omega_R$ determines the oscillation frequency, while the imaginary part $\omega_I$ governs the damping rate. Even more, stability of the black hole requires $\omega_I < 0$, ensuring exponential damping, otherwise the black hole becomes unstable against the type of perturbation applied.
QNMs play a fundamental role in black hole physics, particularly in gravitational wave astronomy and tests of general relativity. Roughly speaking, their study force us to solving complex differential equations, often tackled using not only exact solution but also numerical techniques to find the corresponding QN frequencies. Recent advancements have introduced recurrence-based approaches, improving both accuracy and computational efficiency in QNM frequency calculations.

In recent studies, various aspects of quasinormal modes (QNMs) have been extensively explored, highlighting the rich phenomenology emerging from alternative gravitational theories. Hu and Zhu investigated scalar field perturbations in Einstein-Bumblebee gravity with a cosmological constant, focusing on their QNM spectra and dynamical evolution, thereby uncovering new signatures of Lorentz-violating effects \cite{Hu:2025isj}. Zahid et al. studied rotating charged black holes within Kalb–Ramond gravity, assessing both shadow images and QNMs, and providing insightful comparisons with observational data from the Event Horizon Telescope (EHT), which further constrain alternative theories of gravity \cite{Zahid:2024hyy}. Araújo Filho et al. analyzed the influence of antisymmetric tensor fields on charged black holes, examining phenomena such as evaporation processes, particle geodesics, gravitational lensing, and scattering effects alongside their QNMs, thus illustrating a comprehensive interaction between modified gravity theories and observable astrophysical phenomena \cite{AraujoFilho:2024rcr}. Moreover, Guo, Tan, and Liu addressed Lorentz violation explicitly by calculating QNMs for a specific Lorentz-violating black hole solution, providing additional theoretical foundations to test deviations from classical general relativity through gravitational wave observations \cite{Guo:2023nkd}.

Neutrino pair annihilation, specifically the process \(\nu \bar{\nu} \rightarrow e^- e^+\), is an astrophysically important mechanism extensively studied for its significant role as a potential energy source for gamma-ray bursts (GRBs). GRBs are among the most energetic and luminous events in the cosmos, characterized by sudden, intense emissions of gamma radiation lasting from fractions of a second to several minutes. These bursts are categorized mainly into two types: short-duration GRBs, typically associated with neutron star mergers, and long-duration GRBs, generally linked to core-collapse supernovae of massive stars. Despite considerable observational data, understanding the exact mechanisms powering GRBs remains an active area of research. One leading hypothesis suggests that the annihilation of neutrinos emitted from hot accretion disks around compact stellar objects, such as black holes or neutron stars, substantially contributes to the energy required to power these cosmic phenomena \cite{Popham:1998ab, Janka:1995cq, Ruffert:1996by}.

Recent studies underline the necessity of investigating neutrino pair annihilation within various gravitational theories beyond classical General Relativity (GR). These explorations are critical not only for elucidating the fundamental physics governing compact astrophysical objects but also for utilizing astronomical observations of GRBs to probe new physics in extreme gravitational fields \cite{Lambiase:2020iul, Salmonson:1999es}. Motivated by these considerations, our present work seeks to explore neutrino pair annihilation around gravitational objects characterized by the charged black holes in Kalb-Ramond gravity. Studying neutrino annihilation in the presence of such black hole solution thus offers a valuable opportunity to investigate both theoretical models of LSB and their astrophysical implications, further bridging theoretical predictions with observable phenomena in high-energy astrophysics.

The present study builds upon these foundations by considering electrically charged black holes within a specific Lorentz-violating gravitational theory mediated by a background Kalb-Ramond field. We focus on the coupling of the antisymmetric field to both curvature and the electromagnetic field, exploring how this configuration reshapes the black hole spacetime and influences physical observables such as the black hole shadow, the deflection angle in the weak field regime, quasinormal modes and the energy deposition rate. This framework offers a fertile ground for both theoretical inquiry and observational confrontation, especially in the era of gravitational wave astronomy and black hole imaging.

The work is organized as follows: $G=c=1$, $(-,+,+,+)$.

\section{Brief review of Lorentz-violating gravity with a background KR field and black hole solutions} \label{Model}

In Ref. \cite{}, a modified theory of gravity is introduced that explicitly incorporates spontaneous Lorentz symmetry breaking (LSB) via a non-vanishing vacuum expectation value (VEV) of the Kalb-Ramond (KR) field $ B_{\mu\nu} $, a rank-2 antisymmetric tensor field naturally arising in string theory. Section 2 delineates the foundational action, where the KR field non-minimally couples to curvature through two independent couplings $ \xi_2 $ and $ \xi_3 $, and interacts with the electromagnetic sector via a dimension-4, gauge-invariant term $ \eta B_{\mu\nu} B_{\rho\sigma} F^{\mu\nu} F^{\rho\sigma} $, thus violating local particle Lorentz invariance.

The total action functional considered is:
\begin{equation}
    S = \frac{1}{2} \int d^4x \sqrt{-g} \left[ R - 2\Lambda - \frac{1}{6} H_{\mu\nu\rho}H^{\mu\nu\rho} - V(B_{\mu\nu}B^{\mu\nu} \pm b^2) + \xi_2 B_{\rho\mu} B^\nu_{~\mu} R_{\rho\nu} + \xi_3 B_{\mu\nu}B^{\mu\nu} R \right] + \int d^4x \sqrt{-g} \mathcal{L}_M.
\end{equation}
Here, $ H_{\mu\nu\rho} = \partial_{[\mu} B_{\nu\rho]} $ denotes the field strength of the KR field. To realize charged black hole solutions, the authors introduce a novel matter Lagrangian:
\begin{equation}
    \mathcal{L}_M = -\frac{1}{2} F_{\mu\nu}F^{\mu\nu} - \eta B^{\alpha\beta}B^{\gamma\rho}F_{\alpha\beta}F_{\gamma\rho},
\end{equation}
where $ \eta $ is a coupling parameter mediating the LSB in the gauge sector.

Assuming a static, spherically symmetric spacetime, the metric ansatz is chosen as:
\begin{equation}
    ds^2 = -F(r) dt^2 + G(r) dr^2 + r^2 d\theta^2 + r^2 \sin^2\theta d\phi^2.
\end{equation}
To preserve the vacuum structure that spontaneously breaks Lorentz symmetry, the KR field is fixed to a background configuration with a single non-zero component $ b_{01} = -b_{10} = \tilde{E}(r) $. This configuration ensures a vanishing KR field strength $ H_{\mu\nu\rho} = 0 $, and induces a constant norm condition $ b_{\mu\nu}b^{\mu\nu} = \pm b^2 $.

Crucially, the presence of the interaction term $ \eta B_{\mu\nu} B_{\rho\sigma} F^{\mu\nu} F^{\rho\sigma} $ is necessary to obtain consistent charged solutions. The modified Einstein equations are derived, and the effective stress-energy tensors for both the electromagnetic and KR fields are included. The electromagnetic field equation is modified to:
\begin{equation}
    \nabla^\nu \left( F_{\mu\nu} + 2\eta B_{\mu\nu} B^{\alpha\beta} F_{\alpha\beta} \right) = 0.
\end{equation}

In Section 3, the field equations are solved under the assumption of vanishing cosmological constant $ \Lambda = 0 $. The key simplification arises from the radial equations, yielding the constraint $ F(r)G(r) = 1 $, or equivalently $ G(r) = F^{-1}(r) $. The Maxwell equation yields a Coulomb-like potential:
\begin{equation}
    \Phi(r) = \frac{Q}{(1 - 2\eta b^2)r}.
\end{equation}

A crucial relation between the Lorentz-violating parameter $ \ell = \frac{1}{2}\xi_2 b^2 $ and the interaction coupling $ \eta $ is enforced by consistency of the field equations:
\begin{equation}
    \eta = \frac{\ell}{2b^2}.
\end{equation}

This relation is nontrivial and reveals that the electromagnetic sector is directly sensitive to the spontaneous breaking of Lorentz symmetry. Substituting back, one obtains the electrostatic potential:
\begin{equation}
    \Phi(r) = \frac{Q}{(1 - \ell)r},
\end{equation}
and a modified metric function:
\begin{equation}
    F(r) = \frac{1}{1 - \ell} - \frac{2M}{r} + \frac{Q^2}{(1 - \ell)^2 r^2}.
\end{equation}

This form is structurally equivalent to the Reissner-Nordstr\"om metric, yet with the electric charge rescaled by a factor $ (1 - \ell)^{-1} $, and the entire spacetime modulated by an effective Lorentz-violating factor. The full line element is thus:
\begin{equation} \label{metric}
    ds^2 = -\left( \frac{1}{1 - \ell} - \frac{2M}{r} + \frac{Q^2}{(1 - \ell)^2 r^2} \right) dt^2 + \left( \frac{1}{1 - \ell} - \frac{2M}{r} + \frac{Q^2}{(1 - \ell)^2 r^2} \right)^{-1} dr^2 + r^2 d\theta^2 + r^2 \sin^2\theta d\phi^2.
\end{equation}
This solution, Eq. \eqref{metric} in the text, embodies a Lorentz-violating generalization of the classical RN black hole. The asymptotic limit reveals a spacetime that is not strictly flat, as $ F(r \to \infty) \to \frac{1}{1 - \ell} $, implying an effective vacuum shift. The parameter $ \ell $ must be extremely small to conform to empirical Solar System constraints, yet its imprint on black hole structure is significant, notably altering the location of horizons and the electromagnetic field configuration shown in Tab. \ref{tab:solar_system_limits}.

\begin{table}[h]
\centering
\begin{tabular}{|c|c|}
  \hline
Observation & Allowed Range for \(\ell\) \\
  \hline
Mercury perihelion advance & \(\ell \in [-3.7\times10^{-12},\,1.9\times10^{-11}]\)\\
  \hline
Gravitational light bending & \(\ell \in [-1.1\times10^{-10},\,5.4\times10^{-10}]\)\\
  \hline
Shapiro delay measurement & \(\ell \in [-6.1\times10^{-13},\,2.8\times10^{-14}]\)\\
  \hline
\end{tabular}
\caption{Derived limits on the Lorentz-violating parameter \(\ell\) from Solar System observations \cite{Yang:2023wtu}.}
\label{tab:solar_system_limits}
\end{table}

\section{Deflection Angle}
In the study of black hole gravitational lensing within non-asymptotically flat spacetimes, such as those involving cosmological constants or modified gravity theories, the Gauss-Bonnet theorem (GBT) provides a powerful geometric tool for computing deflection angles of light. The GBT in this context is applied to the optical geometry of the spacetime and is expressed as:
\begin{equation}
    \hat{\alpha} = -\int\int_{\mathcal{D}_R} K \, dS + \int_{\partial \mathcal{D}_R} \kappa_g \, d\ell + \sum_i \theta_i = 2\pi\chi(\mathcal{D}_R),  
\end{equation}
where $ \hat{\alpha} $ is the total deflection angle, $ \mathcal{K} $ is the Gaussian curvature of the optical manifold, $ \kappa_g $ is the geodesic curvature of the boundary $ \partial \mathcal{D}_R $, and $ \chi(\mathcal{D}_R) $ is the Euler characteristic of the domain $ \mathcal{D}_R $. If the domain is now taken from the finite distance of the source and the receiver (from the black hole), bounded by the circular photon orbit \cite{Li:2020wvn}, the above expression can be rewritten as
\begin{equation} \label{wda_Li}
    \hat{\alpha} = \iint_{_{r_{\rm ph}}^{\rm R }\square _{r_{\rm ph}}^{\rm S}}KdS + \phi_{\text{RS}} = \int^{\phi_{\rm R}}_{\phi_{\rm S}} \int_{r_{\rm ph}}^{r(\phi)} K\sqrt{g} \, dr \, d\phi + \phi_{\rm RS},
\end{equation}
where $\phi_{\rm RS} = \phi_{\rm R} - \phi_{\rm S}$, and $\phi_{\rm R} = \pi - \phi_{\rm S}$ are the positional angles. Moreover, $g$ is the determinant of the metric. This formulation respects the finite spatial domain of non-asymptotically flat geometries as it enables the derivation of deflection angles from the topology and intrinsic curvature of the spacetime's optical metric.

For brevity, let us write $A(r) = F(r)$, $B(r) = F(r)^{-1}$, and $C(r) = r^2$. Furthermore, we specialize in equatorial analysis such that $\theta = \pi/2$ and Eq. \eqref{metric} should reduce to
\begin{equation} \label{metric2}
	ds^2 = -A(r) dt^2 + B(r) dr^2 + C(r) d\phi^2
\end{equation}
without loss of generality. The deflection angle can thus be generalized not only for photons but also for massive particles of mass $\mu$. In this case, the Jacobi metric must be used:
\begin{equation} \label{Jac_met}
    dl^2=g_{ij}dx^{i}dx^{j}
    =(E^2-\mu^2A(r))\left(\frac{B(r)}{A(r)}dr^2+\frac{C(r)}{A(r)}d\phi^2\right).
\end{equation}
Here, $E$ is defined as $E = (1-v^2)^{-1/2}$ and set $\mu=1$ for massive particle analysis.

With $K$ and $g$ given as \cite{Pantig:2024kqy}
\begin{equation}
	K=-\frac{1}{\sqrt{g}}\left[\frac{\partial}{\partial r}\left(\frac{\sqrt{g}}{g_{rr}}\Gamma_{r\phi}^{\phi}\right)\right], \qquad g = \frac{(E^2 - \mu^2 A(r))B(r)C(r)}{A(r)^2},
\end{equation}
we can then write
\begin{equation} \label{gct}
	\int_{r_{\rm ph}}^{r(\phi)} K\sqrt{g}dr = -\frac{A(r)\left(E^{2}-A(r)\right)C'-E^{2}C(r)A(r)'}{2A(r)\left(E^{2}-A(r)\right)\sqrt{B(r)C(r)}}\bigg|_{r = r(\phi)},
\end{equation}
which requires the calculation of the orbit equation given by
\begin{equation} \label{orb_eq}
	F(u) \equiv \left(\frac{du}{d\phi}\right)^2 
	= \frac{C(u)^2u^4}{A(u)B(u)}\Bigg[\left(\frac{E}{J}\right)^2-A(u)\left(\frac{1}{J^2}+\frac{1}{C(u)}\right)\Bigg].
\end{equation}
Here, $J$ is defined as $J = vbE$ and is the angular momentum per unit mass, and $b$ is the impact parameter.

In the succeeding calculations, we need some clarifications. First, we retain $E$ and $J$ to reduce clutter. Also useful is writing $\zeta^2 = (1-\ell)^{-1/2}$, and $\tilde{Q}^2 = Q^2(1-\ell)^{-2}$. We have chosen $\zeta^2$ to avoid an ill-defined metric \cite{Majumder:2024mle}. The derivation of the weak deflection angle involves approximations. For instance, if one says that the expression undergoes an approximation of $r \rightarrow \infty$, which is the weak field approximation, the result is identical if we do the approximation $M \sim 0$. Note that this observation is valid in this context due of the structure of the metric function. More often, the approximation $M \sim 0$ simplifies the resulting expression better than doing the $r \rightarrow \infty$ approximation. 

Going back to the orbit equation in Eq. \eqref{orb_eq}, we find
\begin{equation}
	\left(\frac{du}{d\phi}\right)^2 = \frac{E^{2}}{J^{2}}-\left(\tilde{Q}^{2} u^{2}+\zeta^2-2 M u \right) \left(\frac{1}{J^{2}}+u^{2}\right).
\end{equation}
The condition $\left(\frac{du}{d\phi}\right) = 0$, allows one to obtain
\begin{equation}
	u = \frac{1}{b}-\frac{\zeta -1}{v^{2} b}.
\end{equation}
To show the dependence of $u$ to $\phi$ explicitly, we differentiate Eq. \eqref{orb_eq} again, and the condition $\left(\frac{d^2u}{d\phi^2}\right) = 0$ gives a differential equation, with a solution of
\begin{equation}
	u(\phi) = C_1 \sin(\zeta\phi) + C_2 \cos(\zeta\phi).
\end{equation}
Following the procedures presented \cite{Pantig:2024kqy}, and through iterative method, one finds
\begin{equation} \label{e_u(final)}
	u(\phi) = \frac{1}{b}\sin \! \left(\zeta\,\phi \right)+ \frac{M}{b^{2} v^{2}}\left[1 + v^2 \cos \! \left(\zeta\,\phi \right)\right]-\left( \frac{1}{b} + \frac{4m}{b^2}  \right) \frac{\zeta -1}{v^{2}}-\frac{\tilde{Q}^{2}}{2 b^{3} v^{2}}+\frac{\left(4 v^{2}+1\right) \tilde{Q}^{2} \left(\zeta -1\right)}{2 b^{3} v^{4}}.
\end{equation}
Next, using Eq. \eqref{e_u(final)} and going back to Eq. \eqref{gct}, we obtain a quite worked-out equation:
\begin{align}
	&\int_{r_{\rm ph}}^{r(\phi)} K\sqrt{g}dr \sim \frac{\sin \! \left(\phi \right) \left(2 E^{2}-1\right) M}{\left(E^{2}-1\right) b} - \zeta +\frac{\left(-3 E^{2}+1\right) \left(\sin^{2}\left(\phi \right)\right) \tilde{Q}^{2}}{2 E^{2} b^{2}-2 b^{2}} \nonumber \\
	&-\frac{6 \sin \! \left(\phi \right) \left(\zeta -1\right) \tilde{Q}^{2}}{b^{2} v^{2} \left(E^{2}-1\right) \left(2 E^{2}-2\right)}\left[ -\frac{v^{2} \left(E^{4}-\frac{8}{3} E^{2}+\frac{1}{3}\right) \sin \! \left(\phi \right)}{2}+\left(E -1\right) \left(\cos \! \left(\phi \right) \phi  \,v^{2}-1\right) \left(E +1\right) \left(E^{2}-\frac{1}{3}\right) \right] \nonumber \\
	&-\frac{\left(\zeta -1\right) M}{b \,v^{2} \left(E^{2}-1\right) \left(2 E^{2}-2\right)}\left[ -4 \left(E -1\right) \left(E^{2}-\frac{1}{2}\right) v^{2} \left(E +1\right) \phi  \cos \! \left(\phi \right)+4 \left(E^{4}-\frac{5}{2} E^{2}+\frac{1}{2}\right) v^{2} \sin \! \left(\phi \right)+4 E^{4}-6 E^{2}+2 \right] \nonumber \\
	&+ \frac{M \,\tilde{Q}^{2}}{b^{3} v^{2} \left(E^{2}-1\right)^{2}} \left[ \left(E^{2}+1\right) \left(E^{2}+\frac{1}{2}\right) v^{2} \left(\sin^{3}\left(\phi \right)\right)-3 \left(E -1\right) \left(v^{2}+1\right) \left(E +1\right) \left(E^{2}-\frac{1}{3}\right) \sin \! \left(\phi \right)-E^{4}+\frac{3 E^{2}}{2}-\frac{1}{2} \right] \nonumber \\
	& - \mathcal{O}[\left(\zeta -1\right)\tilde{Q}^{2} M].
\end{align}
Next, we integrate the above resulting to:
\begin{align} \label{e_int}
	&\int_{\phi_{\rm S}}^{\phi_{\rm R}} \int_{r_{\rm ph}}^{r(\phi)} K\sqrt{g} \, dr \, d\phi \sim -\frac{\cos \! \left(\phi \right) \left(2 E^{2}-1\right) M}{\left(E^{2}-1\right) b} \bigg\vert_{\phi_{\rm S}}^{\phi_{\rm R}} - \phi_{\rm RS}\left(\zeta -1\right) - \frac{3 \tilde{Q}^{2} \left(E^{2}-\frac{1}{3}\right) \left(-\cos \! \left(\phi \right) \sin \! \left(\phi \right)+\phi \right)}{4 \left(E^{2}-1\right) b^{2}} \bigg\vert_{\phi_{\rm S}}^{\phi_{\rm R}} \nonumber \\
	&+ \frac{3 \left(\zeta -1\right) \tilde{Q}^{2}}{2 b^{2} v^{2} \left(E^{2}-1\right)^{2}} \Bigg\{ \left(E -1\right) \phi  \left(E +1\right) \left(E^{2}-\frac{1}{3}\right) v^{2} \left(\cos^{2}\left(\phi \right)\right)+\left[-\left(E^{4}-2 E^{2}+\frac{1}{3}\right) v^{2} \sin \! \left(\phi \right)-2 E^{4}+\frac{8 E^{2}}{3}-\frac{2}{3}\right] \cos \! \left(\phi \right) \nonumber \\
	&-\frac{2 E^{2} v^{2} \phi}{3}
	\Bigg\}  \bigg\vert_{\phi_{\rm S}}^{\phi_{\rm R}} + \frac{2 \left(\zeta -1\right) M}{b \,v^{2} \left(E^{2}-1\right)^{2}} \left[  2 \left(E^{4}-2 E^{2}+\frac{1}{2}\right) v^{2} \cos \! \left(\phi \right)+\left(E -1\right) \left(E^{2}-\frac{1}{2}\right) \phi  \left(\sin \! \left(\phi \right) v^{2}-1\right) \left(E +1\right) \right] \bigg\vert_{\phi_{\rm S}}^{\phi_{\rm R}} \nonumber \\
	&+ \frac{M \,\tilde{Q}^{2}}{3 b^{3} v^{2} \left(E^{2}-1\right)^{2}} \Bigg\{ \left(E^{2}+1\right) \left(E^{2}+\frac{1}{2}\right) v^{2} \left(\cos^{3}\left(\phi \right)\right)+\left[\left(6 v^{2}+9\right) E^{4}+\left(-\frac{33 v^{2}}{2}-12\right) E^{2}+\frac{3 v^{2}}{2}+3\right] \cos \! \left(\phi \right) \nonumber \\
	& -3 \phi  \,E^{4}+\frac{9 E^{2} \phi}{2}-\frac{3 \phi}{2} \Bigg\} \bigg\vert_{\phi_{\rm S}}^{\phi_{\rm R}} - \mathcal{O}[\left(\zeta -1\right)\tilde{Q}^{2} M].
\end{align}
We now see the necessity to find the expression for $\phi$. Using Eq. \eqref{e_u(final)}, we find
\begin{align}
	&\phi \sim \arcsin \! \left(b u \right)+\frac{ \left(\zeta -1\right)}{v^{2} \sqrt{1-2 b^{2} u^{2}}}+\frac{ \tilde{Q}^{2}}{2 b^{2} v^{2} \sqrt{1- b^{2} u^{2}}}+\frac{\left[v^{2} \left(b^{2} u^{2}-1\right)-1\right] M}{\sqrt{1-b^{2} u^{2}}\, b \,v^{2}} +\frac{u \left(b^{2} v^{2} u^{2}-v^{2}+1\right) \tilde{Q}^{2} M}{\sqrt{1- b^{2} u^{2}}\, \left(2 b^{4} u^{2} v^{4}-2 b^{2} v^{4}\right)} \nonumber \\
	& +\frac{\left(\zeta -1\right)\, \left(b^{3} u^{3} v^{2}+4 b^{2} v^{2} u^{2}+\left(1-v^{2}\right) u b -4 v^{2}\right) M}{\sqrt{1- b^{2} u^{2}}\, v^{4} \left(b^{3} u^{2}-b \right)} -\frac{\left(4 b^{2} v^{2} u^{2}+b^{2} u^{2}+b u -4 v^{2}-1\right) \, \left(\zeta -1\right) \tilde{Q}^{2}}{\sqrt{1- b^{2} u^{2}}\, \left(2 b^{4} u^{2} v^{4}-2 b^{2} v^{4}\right)}  \nonumber \\
	& - \mathcal{O}[\left(\zeta -1\right)\tilde{Q}^{2} M].
\end{align}
Now, by observing Eq. \eqref{e_int}, we need the following expressions for $\cos(\phi)$, and $\cos(\phi)\sin(\phi)$. These are
\begin{align} \label{e_cos}
	&\cos(\phi) \sim \sqrt{1-b^{2} u^{2}}-b u \left(\frac{1}{v^{2} \sqrt{1- b^{2} u^{2}}}-\arcsin \! \left(b u \right)\right) \left(\zeta -1\right)-\frac{u \, \tilde{Q}^{2}}{2 b \,v^{2} \sqrt{1- b^{2} u^{2}}}-\frac{u \left[v^{2} \left(b^{2} u^{2}-1\right)-1\right] M}{\sqrt{1-b^{2} u^{2}}\, v^{2}} \nonumber \\
	& + \mathcal{O}\left[ \left(\zeta -1\right)M, \left(\zeta -1\right)\tilde{Q}^{2}, \tilde{Q}^{2} M, \left(\zeta -1\right)\tilde{Q}^{2} M  \right],
\end{align}
and
\begin{align} \label{e_cos-sine}
	&\cos(\phi)\sin(\phi) \sim b u \sqrt{1-b^{2} u^{2}}+\frac{\left(\arcsin \! \left(b u \right) v^{2} \sqrt{1-b^{2} u^{2}}-1\right) \left(2 b^{2} u^{2}-1\right) \left(\zeta -1\right)}{\sqrt{1- b^{2} u^{2}}\, v^{2}}-\frac{\left(2b^{2} u^{2}-1\right) \, \tilde{Q}^{2}}{2\sqrt{1- b^{2} u^{2}}\, b^{2} v^{2}} \nonumber \\
	& -\frac{2 \left(2b^{2} u^{2}-1\right) \left(b^{2} v^{2} u^{2}-v^{2}-1\right) M}{\sqrt{1-b^{2} u^{2}}\, b \,v^{2}} + \mathcal{O}\left[ \left(\zeta -1\right)M, \left(\zeta -1\right)\tilde{Q}^{2}, \tilde{Q}^{2} M, \left(\zeta -1\right)\tilde{Q}^{2} M  \right],
\end{align}
respectively. We now apply the properties $\cos(\pi-\phi) = \cos(\phi)$, $\sin(\pi-\phi) = \sin(\phi)$, and $\phi_{\rm RS} = \pi - 2\phi$, and also assume that $u_{\rm S} = u_{\rm R} = u$. Using Eq. \eqref{wda_Li}, the implicit expression for the weak deflection angle is
\begin{equation}
	\hat{\alpha}\sim \frac{2 \left(v^{2}+1\right) M}{v^{2} b} \cos(\phi) - \phi  \left(\zeta -1\right) - \frac{\tilde{Q}^{2} \left(v^{2}+2\right)}{4 v^{2} b^{2}} \left( 2 \cos \! \left(\phi \right) \sin \! \left(\phi \right)+\phi \right) + \mathcal{O}\left[ \left(\zeta -1\right)M, \left(\zeta -1\right)\tilde{Q}^{2}, \tilde{Q}^{2} M, \left(\zeta -1\right)\tilde{Q}^{2} M  \right].
\end{equation}
Plugging Eqs. \eqref{e_cos} and \eqref{e_cos-sine} to the above, and with considerable algebraic manipulations, we obtain the general expression for $\hat{\alpha}$
\begin{align}
	&\hat{\alpha} \sim \frac{2 \left(v^{2}+1\right) M}{v^{2} b}\left[ \sqrt{1-b^{2} u^{2}}-b u \left(\frac{1}{v^{2} \sqrt{1- b^{2} u^{2}}}-\arcsin \! \left(b u \right)\right) \left(\zeta -1\right) \right] \nonumber \\
	&-\left( 1-\frac{1}{\zeta} \right) \left[ \pi -2 \arcsin \! \left(b u \right)-\frac{ \tilde{Q}^{2}}{b^{2} v^{2} \sqrt{1- b^{2} u^{2}}}-\frac{2 \left[v^{2} \left(b^{2} u^{2}-1\right)-1\right] M}{\sqrt{1-b^{2} u^{2}}\, b \,v^{2}} \right] \nonumber \\
	&-\frac{\tilde{Q}^{2} \left(v^{2}+2\right)}{4 v^{2} b^{2}} \Bigg[ 2 b u \sqrt{1-b^{2} u^{2}}+\frac{2 \left(\arcsin \! \left(b u \right) v^{2} \sqrt{1- b^{2} u^{2}}-1\right) \left(2 b^{2} u^{2}-1\right) \left(\zeta -1\right)}{\sqrt{1- b^{2} u^{2}}\, v^{2}} \nonumber \\
	& + \frac{1}{\zeta}\left(\pi -2 \arcsin \! \left(b u \right)-\frac{2 \left(\zeta -1\right)}{\sqrt{1- b^{2} u^{2}}\, v^{2}}\right) \Bigg] + \mathcal{O}\left[ \left(\zeta -1\right)M, \left(\zeta -1\right)\tilde{Q}^{2}, \tilde{Q}^{2} M, \left(\zeta -1\right)\tilde{Q}^{2} M  \right].
\end{align}

In the far approximation, where $u \sim 0$, the above equation will reduce to
\begin{align} \label{wda1}
	&\hat{\alpha} \sim \frac{2 \left(v^{2}+1\right) M}{v^{2} \zeta b} -\pi \left( 1-\frac{1}{\zeta} \right) - \frac{\tilde{Q}^{2}}{v^{2} b^{2}} \left[ \frac{\left(\frac{\pi  \,v^{2}}{2}+\left(\zeta -1\right)^{2}\right) \left(v^{2}+2\right)}{2 v^{2} \zeta}+1-\frac{1}{\zeta} \right] + \mathcal{O}\left(\zeta -1\right). \nonumber \\
	& = \frac{2 \left(v^{2}+1\right) M}{v^{2} b}-\frac{Q^{2} \pi  \left(v^{2}+2\right)}{4 v^{2} b^{2}} - \left[ \frac{\pi}{2} + \frac{3 Q^{2} \pi  \left(v^{2}+2\right)}{8 v^{2} b^{2}} + \frac{\left(v^{2}+1\right) M}{v^{2} b} - \frac{Q^{2}}{2 v^{2} b^{2}} \right] \ell.
\end{align}
Eq. \eqref{wda1} demonstrates that the Lorentz-violating parameter $\ell$, sourced by the background Kalb-Ramond field, introduces non-negligible corrections to the standard Reissner-Nordström deflection profile. The expression retains $\ell$-dependent terms even in the limit of vanishing charge, which marks a significant departure from other Lorentz-violating frameworks where such terms often decouple in the null case.

Interestingly, we can approximate the above to also consider slow moving massive particles. That is, when $v \sim 0$, the Eq. \eqref{wda1} reduces to
\begin{equation} \label{wda2}
	\hat{\alpha} \sim \frac{1}{v^2}\left[ \frac{2 M}{b}-\frac{Q^{2} \pi}{2 b^{2}}-\frac{3 Q^{2} \pi  \ell}{4 b^{2}}-\left(-\frac{Q^{2}}{2 b^{2}}+\frac{M}{b}\right) \ell \right] + \frac{2 M}{b}-\frac{Q^{2} \pi}{4 b^{2}}-\frac{3 Q^{2} \pi  \ell}{8 b^{2}}-\left(\frac{\pi}{2}+\frac{M}{b}\right) \ell.
\end{equation}
For ultra-relativistic massive particles, where $v \sim 1$,
\begin{equation} \label{wda3}
	\hat{\alpha} \sim \frac{4 M}{b}-\frac{3 Q^{2} \pi}{4 b^{2}}-\frac{9 Q^{2} \pi  \ell}{8 b^{2}} - \left(\frac{\pi}{2}-\frac{Q^{2}}{2 b^{2}}+\frac{2 M}{b}\right) \ell + \left[ -\frac{4 M}{b}+\frac{Q^{2} \pi}{b^{2}}+\frac{3 Q^{2} \pi  \ell}{2 b^{2}}-\left(\frac{Q^{2}}{b^{2}}-\frac{2 M}{b}\right) \ell \right]\left( v - 1 \right).
\end{equation}
Eq. \eqref{wda2} and Eq. \eqref{wda3} indicate that the deflection angle acquires a velocity-sensitive structure, diverging in the non-relativistic limit due to the $ 1/v^2 $ scaling. This suggests an enhanced deflection signature for slow-moving test particles, opening possibilities for probing LIV effects through massive particle lensing scenarios 

Finally, for the weak field deflection of photons,
\begin{equation} \label{wda_null}
	\hat{\alpha}^{\rm null} \sim \frac{4 M}{b}-\frac{3 Q^{2} \pi}{4 b^{2}} - \left[\frac{\pi}{2}-\frac{Q^{2}}{2 b^{2}}\left(1+\frac{9 \pi}{8}\right)+\frac{2 M}{b}\right] \ell
\end{equation}
further confirms that LIV-induced contributions persist and scale linearly with $\ell$, encoded through both geometric and electromagnetic couplings.

In the regime of weak gravitational fields, one can utilize solar system observations, particularly the deflection of light, to place constraints on parameters appearing in the weak deflection angle formula, such as the one expressed in Eq. \eqref{wda_null}. Within the framework of the parametrized post-Newtonian (PPN) formalism, the light deflection angle is given by \cite{Chen:2023bao}  
\begin{equation} \label{ppn}
    \Theta^{\rm PPN} \backsimeq \frac{4M_{\odot}}{R_{\odot}}\left(\frac{n \pm \Delta}{2} \right),
\end{equation}  
where \( n = 1.9998 \) and \( \Delta = \pm 0.0003 \) \cite{Fomalont_2009}. The parameter \( \Delta \) quantifies the uncertainty in spacetime curvature induced by the Sun’s gravitational field, where \( M_\odot = 1476.61 \, \text{m} \) and \( R_{\odot} = 6.96 \times 10^{8} \, \text{m} \). The angle \( \Theta^{\rm PPN} \), expressed in radians, corresponds to observational data on light deflection around the Sun, particularly measurements obtained via high-precision techniques such as the Very Long Baseline Array (VLBA) \cite{Fomalont_2009}. By comparing Eq. \eqref{ppn} and Eq. \eqref{wda_null}, it becomes possible to derive constraints on the parameter \( \ell \), leading to the relation
\begin{equation} \label{wda_cons}
    \ell \sim -\frac{4 \left(n \pm \Delta -2\right) M_\odot}{R_\odot \pi} - \frac{3 Q_\odot^{2}}{2 R_\odot^{2}} + \frac{Q_\odot^{2} M_\odot}{4 \pi  \,R_\odot^{3}} \left[ \frac{4\left(n \pm \Delta -2\right) \left(9 \pi -4\right)}{\pi}+24 \right].
\end{equation}
Although this expression for \( \ell \) is derived using solar system data, treating the Sun as the gravitational lens, it remains applicable to more massive systems, such as supermassive black holes, provided that the corresponding observational parameter \( (n \pm \Delta) \) is known. 

Recent modeling estimates that the Sun may carry a net electric charge ranging between $1.15 \times 10^{8} \text{ C}$ and $2.80 \times 10^{10} \text{ C}$, based on observations of Earth's orbital mechanics and particle behavior in solar wind \cite{Pei_2025}. Even choosing the highest constraint, the geometrized value of $Q_\odot$ would be very small $(\sim 2.41 \times 10^{-7} \text{ m})$. Using such value, we find that $\ell$ is bounded at $[-2.7013 \times 10^{-10}, 1.3506 \times 10^{-9}]$, coming from the dominance of the first term in eq. \eqref{wda_cons}. These findings not only validate the theoretical consistency of the KR-coupled model in the weak field regime but also establish an observationally viable parameter window for Lorentz-violating effects. The results provide a concrete analytic framework for testing KR-based gravity modifications and position the weak deflection angle as a sensitive observable in future precision lensing experiments.

\section{Shadow analysis}
In this section, we explore the shadow radius analytically, and use EHT data for Sgr. A* and M87* shadow radius to find constraint in the parameter $\ell$. The calculation of the shadow radius is fairly established, and the reader is directed to the works in Refs. \cite{Perlick:2021aok}.

Essential to the calculation of the shadow radius is the location of the photon sphere $r_{\rm ps}$. It can be calculated through the formula
\begin{equation}
    A(r)'r^2 - 2A(r)r = \frac{2 c^{2} r^{2}-6 M r +4 q^{2}}{r} = 0.,
\end{equation}
indicating a signal of the breakdown of photon trapping regions. In a dynamical or quantum context, this could herald instabilities in high-frequency modes, reminiscent of what occurs near extremal black holes where photon spheres degenerate. From the standpoint of observational astrophysics, such regimes may be constrained not only by shadow size but also by the morphology of gravitational wave ringdown, which is also sensitive to photon sphere structure.
The solution that has physical significance is found to be
\begin{equation} \label{rps}
    r_{\rm ps} = \frac{3 M +\sqrt{-8 c^{2} q^{2}+9 M^{2}}}{2 c^{2}}.
\end{equation}
Here, we can see that there is a critical value of $\ell$ for the photonsphere to occur. That is
\begin{equation}
	\ell = 1-\frac{2 \,3^{1/3} Q^{2/3}}{3 M^{2/3}}.
\end{equation}
With Eq. \eqref{rps}, the critical impact parameter is given by
\begin{equation}
    b_\text{crit}^2 = \frac{r_{\rm ps}^2}{A(r_{\rm ps})} = \frac{\left(3 M -8 c^{2} q^{2} +\sqrt{9 M^{2}}\right)^{4}}{8 c^{6} \left(3 M^{2}+M \sqrt{9 M^{2}-8 c^{2} q^{2}}-2 c^{2} q^{2}\right)}.
\end{equation}
According to Ref. \cite{Perlick:2021aok}, the shadow radius can be can be sought off:
\begin{equation}
    R_{\rm sh} = b_{\rm crit}\sqrt{A(r_{\rm obs})} \sim \frac{3 \sqrt{3}\, M}{c^{2}} - \frac{3 \sqrt{3}\, M^{2}}{c^{4} r_{\rm obs}} - \frac{\sqrt{3}\, q^{2}}{2 M} + \frac{\sqrt{3}\, q^{2}}{2 c^{2} r_{\rm obs}} + \mathcal{O}(r_{\rm obs}^{-2})
\end{equation}
In realistic terms, since we aim to find constraints in $\ell$ using EHT results, $M/r_{\rm obs} \sim 0$ and also $Q/r_{\rm obs} \sim 0$. Therefore, the final approximation gives
\begin{equation} \label{eRsh}
	R_{\rm sh} \sim 3 \sqrt{3}\, M \left(3-\frac{2}{\sqrt{1-\ell}}\right) + \frac{2 \sqrt{3}\, Q^{2}}{M}\left(\frac{3}{4}-\frac{1}{\sqrt{1-\ell}}\right).
\end{equation}
Theoretically, Eq. \eqref{eRsh} is a neat equation, especially if $Q$ cannot be neglected. It marks a profound departure from the standard Reissner-Nordström scenario. Here, the Lorentz-violating (LV) parameter $\ell$, originating from the Kalb-Ramond (KR) antisymmetric tensor field, induces a nontrivial deformation to the photon sphere and, subsequently the observable shadow. This dependency on $\ell$ manifests through a multiplicative modulation of both the Schwarzschild term and the charge correction, introducing a composite sensitivity that is simultaneously geometric and electrodynamic. The term proportional to $ Q^2/M $ in Eq. \eqref{eRsh} also merits particular attention. In conventional RN spacetimes, charge reduces the shadow radius due to repulsive effects on photon orbits. However, in this KR-deformed metric, the prefactor $ (3/4 - 1/\sqrt{1 - \ell}) $ introduces a tension: for small but positive $\ell$, the contribution of charge to the shadow radius may become positive—effectively reversing the intuition from Einstein-Maxwell theory. This signals a unique non-minimal coupling between the KR background and the electromagnetic sector, modulated by the coupling parameter $\eta$ (via its relation to $\ell$). This coupling not only re-scales the effective electromagnetic interaction but also alters the causal structure of the effective optical geometry.

The results from the Event Horizon Telescope (EHT) collaboration \cite{EventHorizonTelescope:2019dse,EventHorizonTelescope:2022wkp,EventHorizonTelescope:2021dqv,Vagnozzi:2022moj} are used to find the constraint in $\ell$ using the shadow radius of Sgr A* ($ 4.209M \leq R_{\rm Sch} \leq 5.560M $ at $2\sigma$ level) and M87* ($ 4.313M \leq R_{\rm Sch} \leq 6.079M $ at the $1\sigma$ level). Let the deviation from the standard Schwarzschild case $R_{\rm Sch}$ be $\delta$. Then, for Sgr. A* and M87*, the bounds are $-0.364 \leq \delta/M \leq 0.987$ and $\delta/M = \pm 0.883$, respectively. To further simplify things, there are recent studies pertaining to the value of charges of the black holes in Sgr. A* and M87*. For instance, using observations such as bremsstrahlung surface brightness profiles and X-ray emissions, researchers have placed an observational upper bound of approximately $3 \times 10^{8} \text{ C}$ on the net charge of Sgr A*. It is also mentioned that it can reach a theoretical limit of $\sim 10^{15} \text{ C}$, which is the induced electric charge (Wald charge) due to the black hole rotation while embedded on some magnetic field \cite{Zajacek:2018ycb,Eckart_2019}. This is also the maximum theoretical charge for M87* \cite{EventHorizonTelescope:2021dqv}, and there are certain observational studies that suggest that the charge can also be neglected \cite{EventHorizonTelescope:2021dqv,Zakharov:2021gbg,Banerjee:2019nnj}. With these certain values, if we geometrize the charge $Q$, we obtain around $\sim 10^{-7} \text{ m}$, which suggests that we can safely ignore $Q$ for realistic astrophysical black holes. We then find constraints in $\ell$ using the equation
\begin{equation}
    R_{\rm sh} = R_{\rm Sch} \pm \delta,
\end{equation}
which gives
\begin{equation}
	\ell \sim -\frac{\pm \delta  \sqrt{3}}{9 M}.
\end{equation}
For Sgr. A*, $\ell$ is bounded at $[-0.1899,0.0701]$, while for M87*, $\ell$ is bounded at $[-0.1699,0.1699]$.

\section{QNMs: Scalar and Dirac Perturbations}

In the following, we will discuss how an electrically charged black hole in gravity (with a Kalb-Ramond field) responds to scalar and Dirac perturbations. We will use two complementary approaches: 
 i) the WKB approximation, and 
ii) the Poschl-Teller fitting approach to compute the QNMs given the well-defined behaviour of the effective barrier potential.
%

\subsection{Scalar perturbations}

Let us start with the computation of the quasinormal modes of scalar perturbations. We will denote $\Phi$ as the scalar (and real) perturbation in an electrically charged black hole in gravity (with a Kalb-Ramond field). Now, assuming the action $S[g_{\mu \nu}, \Phi]$, we can derive the standard Klein-Gordon equation \cite{Crispino:2013pya,Kanti:2014dxa,Pappas:2016ovo,Panotopoulos:2019gtn,Avalos:2023ywb,Gonzalez:2022ote,Rincon:2020cos}
\begin{equation}
\frac{1}{\sqrt{-g}}\partial_{\mu}\left(\sqrt{-g}g^{\mu\nu}\partial_{\nu}\Phi\right) = 0.
\end{equation}
Based on the symmetries of the metric and in order to solve the corresponding Klein-Gordon equation for a massless and minimally coupled scalar field, we propose an ansatz in spherical coordinates
\begin{equation}
\Phi(t, r, \theta, \phi) = e^{-i\omega t}\frac{\psi(r)}{r}Y_{l_{b} m}(\theta, \phi).
\end{equation}
As always, $Y_{l_{b} m}(\theta, \phi)$ symbolizes the spherical harmonics, which depend on the angular coordinates. The QN frequency $\omega$ is then obtained by taking appropriate boundary conditions and solving the eigenvalue problem for $\omega$. Doing so, the relevant equation is then written as
\begin{equation} \label{KG}
\frac{\omega^{2}r^{2}}{f(r)} + \frac{r}{\psi(r)}\frac{d}{dr}\left[r^{2}f(r)\frac{d}{dr}\left(\frac{\psi(r)}{r}\right)\right] 
- l_b(l_b + 1) = 0.
\end{equation}
Be aware and notice that in the last equation, we have used the angular part, i.e., 
\begin{equation}
\frac{1}{\sin\theta}\frac{\partial}{\partial\theta}\left(\sin\theta\frac{\partial Y(\Omega)}{\partial\theta}\right) + \frac{1}{\sin^{2}\theta}\frac{\partial^{2}Y(\Omega)}{\partial\phi^{2}} = 
-l_b(l_b + 1)Y(\Omega).
\end{equation}
At this point it should be mentioned that  $l_b(l_b + 1)$ is the eigenvalue, and $l_b$ is the angular degree. 
After mixing the last two equations, it is possible to obtain a second-order differential equation for the radial coordinate. To do this, we introduce the definition of the " tortoise coordinate " $r_{*}$ according to the following expression:
\begin{align}
    r_{*}  \equiv  \int \frac{\mathrm{d}r}{f(r)}\,.
\end{align}
The reduced differential equation for the radial coordinated can be transformed in its Schr{\"o}dinger-like form, namely
\begin{equation} \label{SLE}
\frac{\mathrm{d}^{2}\psi(r_*)}{\mathrm{d}r_{*}^{2}} + \left[\omega^{2} - V(r_*)\right]\psi(r_*) = 0,
\end{equation}
where $V(r)$ is the corresponding effective potential barrier which is defined according to
\begin{equation}\label{poten}
V(r) = f(r)
\Bigg[ 
\frac{l_b(l_b + 1)}{r^{2}} + \frac{f'(r)}{r}
\Bigg].
\end{equation}
Notice that prime represents the derivative with respect to the radial coordinate.
To supplement the wave equation, we need to consider appropriate boundary conditions, which, in this case, are:
\begin{align}
   \Phi \rightarrow \: &\exp(+i \omega r_*), \; \; \; \; \; \;  r_* \rightarrow - \infty ,
   \\
   \Phi \rightarrow \: &\exp(-i \omega r_*), \; \; \; \; \; \; r_* \rightarrow + \infty .
\end{align}
To determine the stability of the black hole solution under scalar perturbations, it is useful at this point to recall that the time dependence of the perturbation is given by $\Phi \sim \exp(-i \omega t)$. Accordingly, a quasinormal frequency with a negative imaginary part corresponds to a decaying mode, which means stability. Conversely, a positive imaginary part indicates a growing mode, which would imply instability in the system.
\begin{figure*}[ht!]
\centering
\includegraphics[scale=0.95]{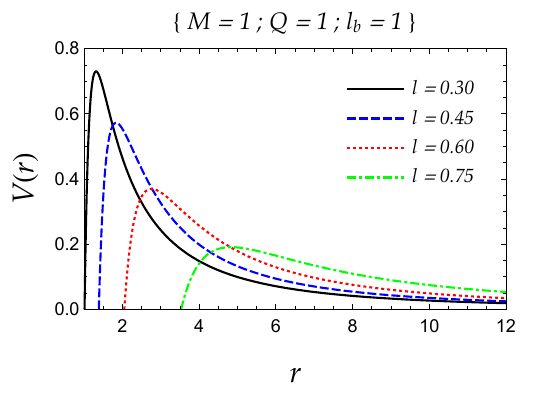} \
\includegraphics[scale=0.95]{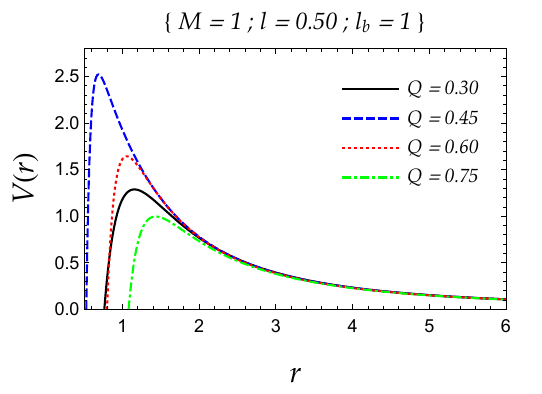} \
\\
\includegraphics[scale=0.95]{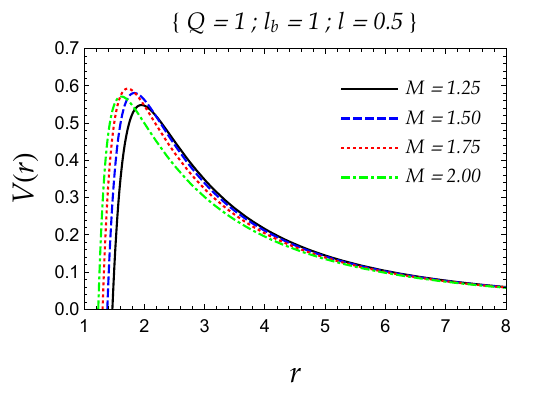} \
\includegraphics[scale=0.95]{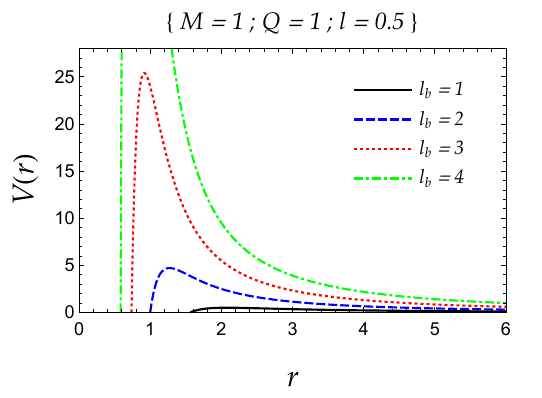} \
\caption{
Effective potential barrier for massless scalar perturbations against the radial coordinate for the parameters shown in the panels. We have four free parameters to vary $\{ \ell , M , Q , l_b \}$. We fixed three of them and vary the remaining one per each sub-figure.
}
\label{fig:1} 	
\end{figure*}
\begin{itemize}
    \item Top Left panel shows the effective potential $V(r)$ versus the radial coordinate for fixed $\{ M , Q , l_b \}$ for different values of the effective parameter $\ell \equiv \xi_2 b^2/2$. 
    It should be noted that as $\ell$ increases, the maximum of the potential continuously decreases and curves shift to the right. At large radii, all solutions converge, as shown in all cases.
    \item Top Right panel shows the effective potential $V(r)$ versus the radial coordinate for fixed $\{ M , \ell , l_b \}$ and different values of the electric charge $Q$. 
    An interesting feature can be seen: as the electric charge increases, the effective potential increases and the maximum shifts to the left. After a certain critical value for the charge, the effective potential starts to decrease and the maximum is now slightly shifted to the right.
    \item Down Left panel shows the effective potential $V(r)$ versus the radial coordinate for fixed $\{ Q , l_b , \ell \}$ and different values of the black hole mass $M$.     In this case we observe that as the mass increases, the effective potential increases and the maximum shifts to the left. After a certain critical value for the charge, the effective potential starts to decrease and the maximum is further shifted to the left.
    \item Down Right panel shows the effective potential $V(r)$ versus the radial coordinate for fixed $\{ M, Q , \ell \}$ and different values of the angular number $l_b$. As we increase $l_b$, the maximum of the effective potential shifts to the left and increases significantly.
\end{itemize}
The behaviour of the effective potential has a peak and decreases asymptotically, making the problem susceptible to the application of the WKB approximation. In order to falsify our results, we will implement an elementary check using the Poschl-Teller fitting approach.
%
\subsection{Dirac Perturbations}
In this subsection we collect the basic expressions for the calculation of quasinormal modes of neutral Dirac particles. We start by assuming a spherically symmetric background in a 4-dimensional spacetime (for more details on the general case see \cite{Cho:2007zi}).
The line element is given by 
\begin{align}
    \mathrm{d} s^{2}=-\left|g_{t t}\right| \mathrm{d} t^{2}+g_{r r} d r^{2} + r^{2}\mathrm{d}\Omega^{2}_{2}, 
\end{align}
where $d\Omega^{2}_{2}$ is the metric for the $(2)$-dimensional sphere and the two metric potentials are then given as 
\begin{align}
g_{tt} &= \frac{1}{1-\ell}\!-\!\frac{2M}{r}\!+\!\fc{Q^2}{\lt(1-\ell \rt)^2r^2}
\equiv 
f(r) =\frac{1}{g_{rr}}, 
\end{align}
Making a conformal transformation over the fields $g_{\mu \nu}$ and $\psi$ (see \cite{Das:1996we,Gibbons:1993hg} and references therein), we have:
\begin{eqnarray}
g_{\mu\nu} & \rightarrow & \overline{g}_{\mu\nu}=\Omega^{2}g_{\mu\nu} , \\
\psi & \rightarrow & \overline{\psi}=\Omega^{-3/2}\psi , \\
\gamma^{\mu}\nabla_{\mu}\psi & \rightarrow & \Omega^{5/2} \overline{\gamma}^{\mu}\overline{\nabla}_{\mu}\overline{\psi} ,
\end{eqnarray}
Now, the metric becomes, after considering $\Omega=1/r$
\begin{align}
\mathrm{d}\overline{s}^{2} = -\frac{1}{r^{2}}f(r)\mathrm{d}t^{2} + \frac{1}{r^{2}} 
f(r)^{-1}\mathrm{d}r^{2} + \mathrm{d}\Omega^{2}_{2} , 
\end{align}
and where the new field is related to the original one as follow $\overline{\psi}=r^{3/2}\psi$.
We can separate the $t$-$r$ sector from the angular part on the 2-sphere, which allows us to decouple the Dirac equation and focus on solving the radial-temporal part. Taking this into account, the Dirac equation for massless fermions can be written in the form
\begin{align}
\overline{\gamma}^{\mu}\overline{\nabla}_{\mu}\overline{\psi}  =  
\bigg[ 
\left(
\overline{\gamma}^{t}\overline{\nabla}_{t} +
\overline{\gamma}^{r}\overline{\nabla}_{r} \right) \otimes 1
+ 
\overline{\gamma}^{5} \otimes
\left( \overline{\gamma}^{a}
\overline{\nabla}_{a}\right)_{S_{2}} 
\bigg] 
\overline{\psi} 
= 0 , 
\end{align}
with $(\overline{\gamma}^{5})^{2}=1$. 
Let us change our notation by ignoring the bars for simplicity.
Let $\chi_{l_b}^{(\pm)}$ be the eigenspinors for the $2$-sphere:
\begin{equation}
\left( \gamma^{a}\nabla_{a} \right)_{S_{2}}\chi_{l_b}^{(\pm)} = \pm i \left( l_b + 1\right) \chi_{l_b}^{(\pm)} ,
\end{equation}
with $l_b = 0, 1, 2, \dots$. Also, it is well-known that the eigenspinors are orthogonal, the reason why it is possible to expand $\psi$ as:
\begin{equation}
\psi = \sum_{l_b} \left( \phi_{l_b}^{(+)} \chi_{l_b}^{(+)} + \phi_{l_b}^{(-)} \chi_{l_b}^{(-)} \right) .
\end{equation}
After replacing, the Dirac equation can be written as:
\begin{equation}\label{eqn:2Ddirac}
\left \{ 
\gamma^{t} \nabla_{t} + \gamma^{r} \nabla_{r} + \gamma^{5} \left[ \pm i \left( l_b + 1 \right) \right] \right \} 
\phi_{l_b}^{(\pm)} = 0 ,
\end{equation}
to obtain the corresponding $2$-dimensional Dirac equation. The solution can be obtained by explicitly choosing the Dirac matrices, i.e.:
\begin{align}
\gamma^{t} &= \frac{r}{\sqrt{f(r)}}(-i\sigma^{3})
\hspace{1cm}
, 
\hspace{1cm}
\gamma^{r} = \sqrt{f(r)} 
\ r\sigma^{2} .
\end{align}
Let us remind that $\sigma^{i}$ are the so-called Pauli matrices, defined in the usual way:
\begin{equation}
\sigma^{1}=\left(
\begin{array}{cc}
0 & 1 \\ 1 & 0
\end{array}
\right)\ \ \ ,\ \ \ \sigma^{2}=\left(
\begin{array}{cc}
0 & -i \\ i & 0
\end{array}
\right)\ \ \ ,\ \ \ \sigma^{3}=\left(
\begin{array}{cc}
1 & 0 \\ 0 & -1
\end{array}
\right) .
\end{equation}
Also, $\gamma^{5}$ is written in term of the Pauli matrices via the expression $\gamma^{5} = (-i\sigma^{3})(\sigma^{2}) = - \sigma^{1}$.
The spin connections are computed and they are:
\begin{align}
\Gamma_{t} &= \sigma^{1} 
\left( 
\frac{1}{4} r^{2} 
\right) 
\frac{\mathrm{d}}{\mathrm{d}r} \left( \frac{f(r)}{r^{2}} \right) 
\hspace{1cm}
, 
\hspace{1cm}
\Gamma_{r} = 0 .
\end{align}
It should be noted that we have two alternatives, i) positive sign and ii) negative sign. We will restrict ourselves to the positive sign. We can rewrite the angular number $l_b$ in terms of an effective parameter $\xi$, which varies depending on the dimension. In four dimensions, the parameter has the simple form
\begin{align}
    \xi \equiv l_b + 1, 
\end{align}
and then $\xi$ takes the values $ +1, +2,...$. Replacing into the Dirac equation, we arrive to the simplified equation
\begin{align}
& \left\{ \frac{r}{\sqrt{f(r)}}(-i\sigma^{3}) \left[
\frac{\partial}{\partial t} + 
\sigma^{1} 
\left( 
\frac{1}{4} r^{2} 
\right) 
\frac{\mathrm{d}}{\mathrm{d}r} \left( \frac{f(r)}{r^{2}} \right)
\right] + 
\sqrt{f(r)} \ 
r \sigma^{2}
\frac{\partial}{\partial r} + (-\sigma^{1})(i) \left( \xi \right) \right\} \phi_{l_b}^{(+)} = 0  
\end{align}
Finally, we split the spatial and temporal part to obtain a first order partial differential equation for $\phi_{l_b}^{(+)}$, i.e., 
\begin{align}
\sigma^{2} 
\left( 
 \sqrt{f(r)} \
r
\right) 
\left[ \frac{\partial}{\partial r} + \frac{r}{2\sqrt{f(r)}}
\frac{\mathrm{d}}{\mathrm{d}r} \left( \frac{\sqrt{f(r)}}{r} \right) \right]
\phi_{l_b}^{(+)} - i \sigma^{1} \xi 
\phi_{l_b}^{(+)} = i \sigma^{3} \left( \frac{r}{\sqrt{f(r)}} \right)
\frac{\partial \phi_{l_b}^{(+)}}{\partial t} . &
\end{align}
The solutions can be obtained replacing $\phi_{l_b}^{(+)}$ in term of the variables $\{ G(r) , F(r) \}$
\begin{equation}
\phi_{l_b}^{(+)} = \left( \frac{\sqrt{f(r)}}{r} \right)^{-1/2} e^{-i \omega t} \left(
\begin{array}{c}
iG(r) \\ F(r)
\end{array}
\right) ,
\end{equation}
Using the last change, the Dirac equation is:
\begin{equation}
\sigma^{2} \left( \sqrt{f(r)} \ r \right) \left(
\begin{array}{c}
i\frac{\mathrm{d}G(r)}{\mathrm{d}r} \\ 
\ \frac{\mathrm{d}F(r)}{\mathrm{d}r}
\end{array}
\right) -i \sigma^{1} \xi \left(
\begin{array}{c}
i G(r) 
\\ 
F(r)
\end{array}
\right) = \sigma^{3} \omega \left( \frac{r}{\sqrt{f(r)}} \right) \left(
\begin{array}{c}
i G(r) 
\\ 
F(r)
\end{array}
\right) .
\end{equation}
After manipulate the Dirac equation and separate each component we get two coupled first-order differential equations for $G \equiv G(r)$ and $F \equiv F(r)$ as follows:
\begin{eqnarray}
f(r) \frac{\mathrm{d}G(r)}{\mathrm{d}r} 
- 
\Bigg [ \frac{\sqrt{f(r)}}{r} \xi \Bigg ] G(r) & = & + \omega F(r) ,  \label{Eq1}
\\ 
f(r) \frac{\mathrm{d}F(r)}{\mathrm{d}r} 
+ 
\Bigg [ \frac{\sqrt{f(r)}}{r} \xi \Bigg ] F(r) & = & - \omega G(r) . \label{Eq2}
\end{eqnarray}
To rewrite the last set of equations, we should take advantage of two useful definitions: i) the  tortoise coordinate, $r_*$, and ii) the auxiliary function $W(r)$, defined as follows
\begin{align}
r_*(r) &\equiv \int ^{r}\mathrm{d}\bar{r} 
\Bigg[
\frac{1}{f(\bar{r})}
\Bigg],
\\
W(r) &\equiv \frac{\xi \sqrt{f(r)}}{r}.
\end{align}
After the appropriate substitutions, the set of first-order linear coupled equations \eqref{Eq1} and \eqref{Eq2} can be reduced to
\begin{align}
    \Bigg[
    \frac{\mathrm{d}}{\mathrm{d}r_{*}} - W(r)
    \Bigg] G(r) &= + \omega F(r)
    \\
    \Bigg[
    \frac{\mathrm{d}}{\mathrm{d}r_{*}} + W(r)
    \Bigg] F(r) &= - \omega G(r)
\end{align}
The coupled first-order differential equations for $G(r)$ and $F(r)$ can be simply decoupled, resulting in two independent Schrodinger-like equations with corresponding effective potentials, namely
\begin{align}
\frac{\mathrm{d}^2F}{\mathrm{d}{r_{*}}^2} + [\omega^2 - V_{-}] F & =  0 , \label{SL1}
\\
\frac{\mathrm{d}^2G}{\mathrm{d}{r_{*}}^2} + [\omega^2 - V_{+}] G & =  0 , \label{SL2}
\end{align}
and the effective potentials are then
\begin{equation}
V_{\pm} = W^2 \pm \frac{\mathrm{d}W}{\mathrm{d}r_{*}} .
\end{equation}
Since the black hole background under consideration fulfills the Schwarzschild ansatz, the tortoise coordinate takes a simplified form that depends only on the function $f(r)$. The treatment of the problem is basically the same regardless of the potentials $V_{+}$ and $V_{-}$. That is, the associated functions $F$ and $G$ satisfy the same differential equation, and although we focus on $\phi_{l_b}^{(+)}$, the approach is the same for $\phi_{l_b}^{(-)}$.
Having established these properties, we now consider the positive parity sector. In terms of the radial coordinate, the effective potential has the following form
\begin{align}
V_{+} &= f(r) 
\Bigg[
\frac{\xi ^2}{r^2} + 
\Bigg(
\frac{\xi  f'(r)}{2 r \sqrt{f(r)}}-\frac{\xi  \sqrt{f(r)}}{r^2}
\Bigg)
\Bigg]
\end{align}
and after introducing the concrete form of the lapse function, we find the effective potential explicitly.
As in the scalar case, some boundary conditions are required to close the problem. Thus, nothing should come in from asymptotic infinity to disturb the system, and nothing should come out from the horizon, which obliges us to consider the following boundary conditions for Schrodinger-like equations \eqref{SL1} and \eqref{SL2}, which are
\begin{align}
   \{F, G\} \rightarrow \: &\exp(+i \omega r_*), \; \; \; \; \; \;  r_* \rightarrow - \infty ,
   \\
   \{F, G\} \rightarrow \: &\exp(-i \omega r_*), \; \; \; \; \; \; r_* \rightarrow + \infty .
\end{align}
At this point, we should consider that, as in the case of scalar perturbations, a negative imaginary part implies a decaying (stable) mode according to our definition $\psi \sim \exp(-i \omega t)$. Of course, the opposite is also true, i.e., we get an unstable mode when $\omega_I >0$.
To illustrate the behaviour of the effective potential with respect to the radial coordinate, we show in figures how the potential looks when the set of parameters $\{ \ell , M , Q , \xi \}$ is varied.

\begin{figure*}[ht!]
\centering
\includegraphics[scale=0.95]{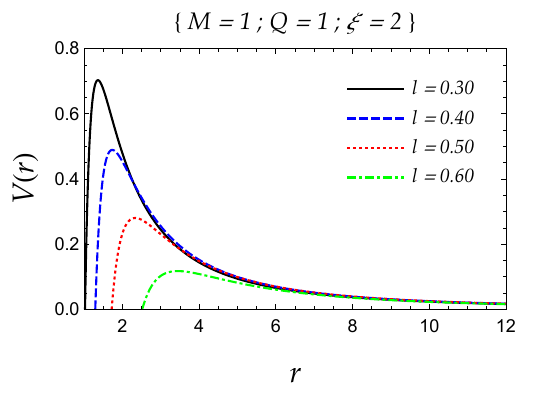} \
\includegraphics[scale=0.95]{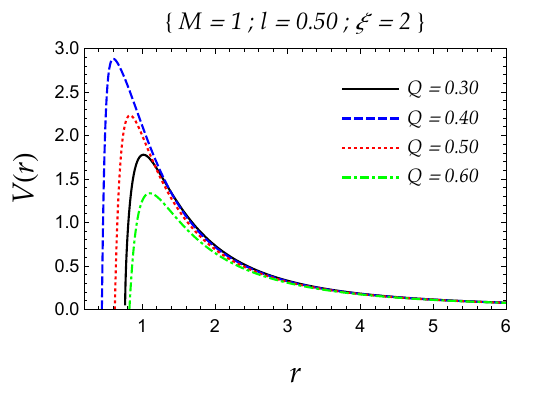} \
\\
\includegraphics[scale=0.95]{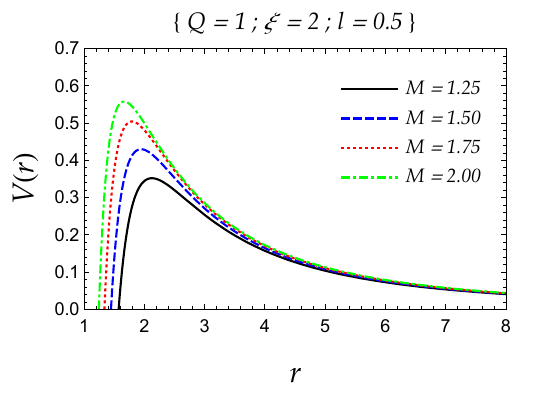} \
\includegraphics[scale=0.95]{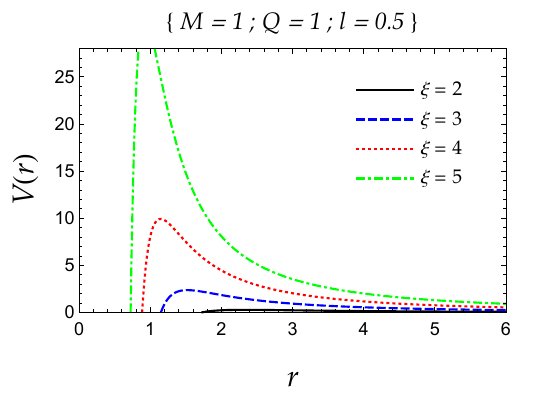} \
\caption{
Effective potential barrier for massless Dirac perturbations against the radial coordinate for the parameters shown in the panels. We have four free parameters to vary $\{ M , Q , \ell , \xi \}$. We fixed three of them and vary the remaining one per each sub-figure.
}
\label{fig:2} 	
\end{figure*}

In particular, from Fig.~\eqref{fig:2}, we clearly identify the following:
\begin{itemize}
    \item Top-Left panel shows the $V(r)$ for fixed $\{ \xi , Q , M \}$ and different values of the effective value $\ell$.
    The effective potential reaches a maximum which decreases as $\ell$ increases. The maximum is also shifted to the right as $\ell$ increases.
    \item Top-Right panel shows the $V(r)$ for fixed $\{ M , \ell , \xi \}$ and different values of the electric charge $Q$. 
    In particular, it is interesting to note that as we increase $Q$, the maximum of the potential increases and eventually decreases. 
    \item Bottom-Left panel shows the $V(r)$ for fixed $\{ Q , \xi , \ell \}$ and different values of the mass $M$.
    It is clear that when the black hole mass increases, the potential is well-defined, the maximum also increases and the curve shift to the left.
    \item Bottom-Right panel shows the $V(r)$ for fixed $\{ M, Q, \ell \}$ and different values of the parameter $\xi$.
    The figure indicates that as $\xi$ increases, so do the effective potential and the maximum shift to the left.
\end{itemize}
As the figures confirm, the potential has a well-defined form and decreases asymptotically. Thus, one of the more convenient methods to obtain the QN frequencies is the WKB method. Moreover, we will concentrate on the third-order QNMs in order to obtain as accurate results as possible.

\subsection{WKB approximation}


The Wentzel-Kramers-Brillouin (WKB) semi-classical approximation is a robust and extensively utilized method for calculating quasinormal mode (QNM) frequencies. Its applicability in this research is particularly justified due to the smoothly varying and regular characteristics of the effective potential, a fact evident in relevant graphical representations (refer to \cite{Schutz:1985km,Iyer:1986np,Iyer:1986nq,Kokkotas:1988fm,Seidel:1989bp} for comprehensive details).

The method has progressively evolved and improved over time. Initially introduced to first order by Schutz and Will \cite{Schutz:1985km}, it was subsequently developed to second and third order accuracy by Iyer and Will \cite{Iyer:1986np}. Later, Konoplya significantly advanced the approach, extending it to sixth order \cite{Konoplya:2003ii}. Further refinements were made by Matyjasek and Opala, who successfully elevated the WKB method's accuracy up to the 13th order \cite{Matyjasek:2017psv}, considerably enhancing its precision and versatility.

The WKB approximation, especially in its lower-order forms, remains particularly effective for determining the fundamental modes of black holes, particularly in Schwarzschild-type geometries. Generally, its accuracy is positively correlated with higher angular harmonic numbers ($l_b$, proportional to $\xi$), but it diminishes with increasing overtone number.

Additionally, the WKB method has been expanded to more complex scenarios, including cases involving three turning points associated with massive scalar fields \cite{Galtsov:1991nwq}. Simone and Will first implemented the WKB approximation for massive scalar perturbations in Schwarzschild and Kerr geometries \cite{Simone:1991wn}. Furthermore, this approach has proven effective in calculating reflection and transmission coefficients for various scattering problems across diverse gravitational backgrounds, employing higher-order corrections (for further discussion, see Ref. \cite{Konoplya:2010vz}).

Numerous studies have successfully applied the WKB technique, refined with higher-order corrections, to evaluate QNMs of black holes and other compact astrophysical objects. A broad range of publications illustrates the technique’s wide applicability and effectiveness (see, for instance, \cite{Konoplya:2023ahd,Konoplya:2020bxa,Konoplya:2003dd,Rincon:2025buq,Balart:2024rtj,EslamPanah:2024qlu,Rincon:2024won,Balart:2023odm,Rincon:2023hvd,Balart:2023swp,Panotopoulos:2019qjk,Panotopoulos:2016wuu}).

Fundamentally, the WKB approximation involves solving a one-dimensional Schrödinger-like equation characterized by a potential barrier. The method matches asymptotic solutions expanded in a Taylor series around the potential peak located at $x = x_0$. This expansion holds between the classical turning points, which are determined by the condition $U(x, \omega) \equiv V(x) - \omega^2 = 0$.

In our investigation, we adopt the WKB method (at different orders) to determine the QNM frequencies of black holes. The general expression for these frequencies is
\begin{equation}
\omega_n^2 = V_0 + \sqrt{-2V_0''}\, \Lambda(n) - i\, \nu \sqrt{-2V_0''} \left[1 + \Omega(n)\right], 
\end{equation}
where the parameters are explicitly defined as follows:
(i) $V_0$ represents the effective potential's maximum at its peak;
(ii) $V_0''$ is the second derivative of the potential at this peak, indicating the curvature;
(iii) $\nu = n + 1/2$, with $n = 0, 1, 2, \dots$ designating overtone modes related to damping;
(iv) The terms $\Lambda(n)$ and $\Omega(n)$ are correction functions accounting for deviations beyond the initial WKB approximation, detailed in \cite{Kokkotas:1988fm} and refined in \cite{Konoplya:2019hlu}.

We implemented the WKB method computationally using \textit{Wolfram Mathematica}, enabling the calculation of quasinormal frequencies accurately up to the 13th order \cite{wolfram,Konoplya:2019hlu,Hatsuda:2019eoj}. This approach is particularly beneficial for systematically analyzing the behavior of QNMs in various black hole geometries, provided the effective potential is single-peaked and sufficiently smooth.

Our analysis focuses on conditions where the overtone number $n$ remains smaller than the angular harmonic number $\xi$ (or equivalently, $\ell$), i.e., $n < \xi$. In such regimes, the WKB approximation provides dependable and precise results. It is well-established that the accuracy of this approximation increases with larger angular harmonic numbers but tends to degrade at higher overtone numbers, thus validating our chosen limitations.

Finally, we have computed the quasinormal modes (QNMs) using the WKB approximation for both scalar (10th order) and Dirac (3rd order) perturbations, with the results summarized in Tables~\eqref{tab:1} and~\eqref{tab:2}. According to our analysis, the black hole is stable under both types of perturbations, given that the imaginary part of the quasinormal frequencies, $\omega_I$, is consistently negative, indicating that all modes are damped and no instabilities appear.

\subsection{Poschl-Teller approximation method}

As a confirmation check, and to gain analytical insight into the quasinormal mode spectrum of black holes, a useful approach is to approximate the effective potential near its maximum by the well-known Poschl-Teller potential. This method is quite good in the eikonal limit, where the angular momentum number $\ell$ is large and the effective potential has a sharp peak. The Poschl-Teller potential is given by 
\begin{align}
V(x) &= \frac{V_0}{\cosh^2[\alpha(x - x_0)]}.
\end{align} 
The parameters involved are: i) $V_0$, which is the height of the potential barrier at its maximum $x = x_0$, and ii) $\alpha$, which controls the width and curvature of the barrier. 
This form allows an exact analytical solution of the wave equation and thus serves as a valuable approximation to the true black hole potential near its maximum.

The quasinormal frequencies associated with this potential can be accurately calculated and are given by the expression
\begin{align}
\omega_n &= \sqrt{V_0 - \frac{1}{4}\alpha^2} - i\,\alpha \left(n + \frac{1}{2}\right), \quad n = 0, 1, 2, \dots
\end{align}

To apply this method in the black hole context, one begins by locating the maximum of the effective potential, denoted by $x_0$, and evaluating both the potential height $V_0 = V(x_0)$ and its second derivative $V''(x_0)$ with respect to the tortoise coordinate. The curvature parameter $\alpha$ is then determined by the relation $\alpha^2 = -V''(x_0)/(2V_0)$, which defines the width of the potential barrier. With these quantities, the Poschl-Teller formula yields quasinormal frequencies that accurately approximate the fundamental modes, especially for low overtones and large angular momentum $\ell$.

Although not as powerful as fully numerical or higher-order semi-analytic methods such as the WKB approximation, the Poschl-Teller approach provides valuable analytical control and physical transparency. It is particularly useful for comparison with more sophisticated techniques and for verification of numerical implementations. Furthermore, the method emphasises the central role of the shape of the effective potential near its peak in determining the QNM spectrum, making it a pedagogically and computationally attractive tool for investigating perturbations in various black hole spacetimes.

Finally, we have used the Poschl-Teller method to cross-check our numerical results obtained via the WKB approximation, finding excellent agreement between the two approaches. Using this method, we have computed the quasinormal modes (QNMs) for both scalar and Dirac perturbations, with the corresponding results summarized in Tables~\eqref{tab:1B} and~\eqref{tab:2B}. As is well known, in the Poschl-Teller approximation, the overtone number $n$ mainly affects the imaginary part of the frequency. Consequently, as $n$ increases, only $\omega_I$ (the imaginary part) is modified. Our results indicate that the black hole is stable under both scalar and Dirac perturbations, as evidenced by the negative sign of the imaginary part of the QNM frequencies.

\begin{figure*}[ht!]
\centering
\includegraphics[scale=0.95]{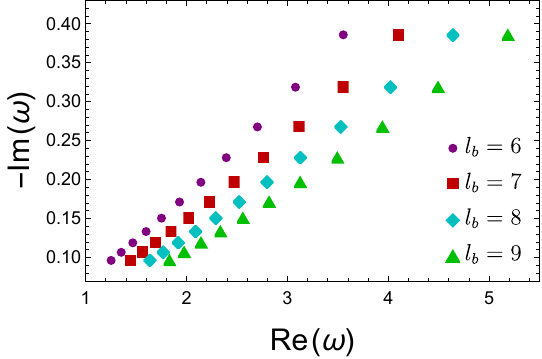} \
\includegraphics[scale=0.95]{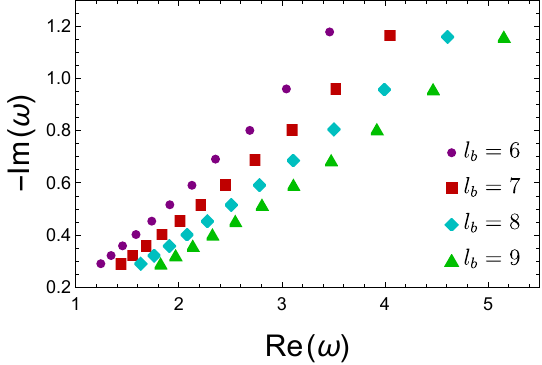} \
\caption{
Quasinormal modes for massless scalar perturbations.
{\bf{Left Panel:}} Quasinormal modes for $M=1$, $Q=1/50$ varying $\ell$ from 0.0 to 0.5 for different values of the parameter $l_b$ for the fundamental mode $(n=0)$.
{\bf{Right Panel:}} Quasinormal modes for $M=1$, $Q=1/50$ varying $\ell$ from 0.0 to 0.5 for different values of the parameter $l_b$ for the first exited mode $(n=1)$.}
\label{fig:3} 	
\end{figure*}
\begin{figure*}[ht!]
\centering
\includegraphics[scale=0.95]{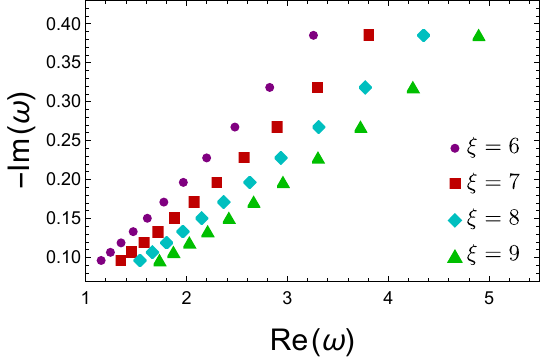} \
\includegraphics[scale=0.95]{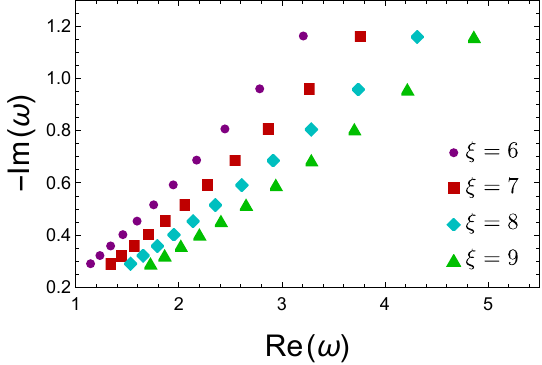} \
\caption{
Quasinormal modes for massless Dirac perturbations.
{\bf{Left Panel:}} Quasinormal modes for $M=1$, $Q=1/50$ varying $\ell$ from 0.0 to 0.5 for different values of the parameter $\xi$ for the fundamental mode $(n=0)$.
{\bf{Right Panel:}} Quasinormal modes for $M=1$, $Q=1/50$ varying $\ell$ from 0.0 to 0.5 for different values of the parameter $\xi$ for the first exited mode $(n=1)$.}
\label{fig:4} 	
\end{figure*}

\begin{table*}
    \centering
	\caption{Quasinormal frequencies of massless scalar perturbations computed using the 10th-order WKB approximation for $M = 1$ and $Q = 1/50$, with varying values of $\{\ell, n, l_b\}$.}
	\label{tab:1}       
	\begin{tabular}{c|c|c|c|c|c|c}
		\hline\noalign{\smallskip}
		$\ell$  & $n$ & $l_b=6$ & $l_b=7$ & $l_b=8$ & $l_b=9$ & $l_b=10$  \\
		\noalign{\smallskip}\hline\noalign{\smallskip}
          \noalign{\smallskip}\hline
        0.00 & 0 & 1.25197\, -0.0963068 i & 1.44429\, -0.096288 i & 1.63667\, -0.0962741 i & 1.82906\, -0.0962647 i & 2.02145\, -0.0962579 i \\
        0.00 & 1 & 1.24382\, -0.2897420 i & 1.43703\, -0.289522 i & 1.63038\, -0.2893080 i & 1.82346\, -0.2891790 i & 2.01638\, -0.2890900 i \\
        0.00 & 2 & 1.22728\, -0.4858490 i & 1.42136\, -0.485197 i & 1.61770\, -0.4838650 i & 1.81242\, -0.4832380 i & 2.00625\, -0.4828770 i \\
        \noalign{\smallskip}\hline    
        0.05 & 0 & 1.35238\, -0.1067180 i & 1.56005\, -0.106694 i & 1.76779\, -0.1066780 i & 1.97554\, -0.1066670 i & 2.18331\, -0.1066590 i \\
        0.05 & 1 & 1.34319\, -0.3211200 i & 1.55180\, -0.320842 i & 1.76064\, -0.3205990 i & 1.96918\, -0.3204510 i & 2.17755\, -0.3203460 i \\
        0.05 & 2 & 1.32590\, -0.5382870 i & 1.53430\, -0.537661 i & 1.74620\, -0.5362930 i & 1.95665\, -0.5355740 i & 2.16610\, -0.5351340 i \\   
		\noalign{\smallskip}\hline
        0.10 & 0 & 1.46687\, -0.1189180 i & 1.69216\, -0.118882 i & 1.91741\, -0.1188640 i & 2.14269\, -0.1188510 i & 2.36799\, -0.1188420 i \\
        0.10 & 1 & 1.45460\, -0.3583140 i & 1.68309\, -0.357470 i & 1.90947\, -0.3572200 i & 2.13545\, -0.3570750 i & 2.36139\, -0.3569600 i \\
        0.10 & 2 & 1.41912\, -0.6073620 i & 1.66628\, -0.598168 i & 1.89533\, -0.5969940 i & 2.12147\, -0.5967710 i & 2.34831\, -0.5963840 i \\
		\noalign{\smallskip}\hline
        0.15 & 0 & 1.59864\, -0.1333250 i & 1.84403\, -0.133284 i & 2.08938\, -0.1332650 i & 2.33480\, -0.1332490 i & 2.58026\, -0.1332380 i \\
        0.15 & 1 & 1.58494\, -0.4017070 i & 1.83381\, -0.400772 i & 2.08005\, -0.4005710 i & 2.32639\, -0.4003790 i & 2.57267\, -0.4002260 i \\
        0.15 & 2 & 1.54606\, -0.6806830 i & 1.81636\, -0.670018 i & 2.06198\, -0.6700960 i & 2.30963\, -0.6694390 i & 2.55771\, -0.6687150 i \\
		\noalign{\smallskip}\hline
        0.20 & 0 & 1.75145\, -0.1505120 i & 2.02004\, -0.150476 i & 2.28871\, -0.1504490 i & 2.55746\, -0.1504310 i & 2.82625\, -0.1504180 i \\
        0.20 & 1 & 1.73791\, -0.4529740 i & 2.00814\, -0.452560 i & 2.27782\, -0.4522810 i & 2.54780\, -0.4520310 i & 2.81737\, -0.4518760 i \\
        0.20 & 2 & 1.71630\, -0.7576000 i & 1.98826\, -0.756826 i & 2.25647\, -0.7568290 i & 2.52944\, -0.7556330 i & 2.79944\, -0.7552670 i \\
        \noalign{\smallskip}\hline
        0.25 & 0 & 1.93006\, -0.1712730 i & 2.22594\, -0.171216 i & 2.52186\, -0.1711860 i & 2.81792\, -0.1711640 i & 3.11400\, -0.1711490 i \\
        0.25 & 1 & 1.91303\, -0.5159540 i & 2.21179\, -0.515022 i & 2.50873\, -0.5147430 i & 2.80659\, -0.5143820 i & 3.10362\, -0.5141950 i \\
        0.25 & 2 & 1.88033\, -0.8674090 i & 2.18689\, -0.861745 i & 2.47991\, -0.8627040 i & 2.78526\, -0.8599490 i & 3.08323\, -0.8594610 i \\
	\noalign{\smallskip}\hline
        0.30 & 0 & 2.14168\, -0.1965930 i & 2.46931\, -0.196570 i & 2.79755\, -0.1965250 i & 3.12577\, -0.1965000 i & 3.45413\, -0.1964800 i \\
        0.30 & 1 & 2.12765\, -0.5905080 i & 2.45069\, -0.591830 i & 2.78271\, -0.5908980 i & 3.11188\, -0.5906770 i & 3.44187\, -0.5903490 i \\
        0.30 & 2 & 2.14059\, -0.9674730 i & 2.40317\, -0.997878 i & 2.75574\, -0.9884100 i & 3.08236\, -0.9890550 i & 3.41826\, -0.9867700 i \\
		\noalign{\smallskip}\hline
        0.35 & 0 & 2.39353\, -0.2281070 i & 2.76064\, -0.227992 i & 3.12736\, -0.2279400 i & 3.49416\, -0.2279060 i & 3.86105\, -0.2278810 i \\
        0.35 & 1 & 2.35613\, -0.6909280 i & 2.73897\, -0.686466 i & 3.10941\, -0.6855170 i & 3.47808\, -0.6850670 i & 3.84644\, -0.6847490 i \\
        0.35 & 2 & 2.18662\, -1.2234800 i & 2.68877\, -1.155710 i & 3.07724\, -1.1471300 i & 3.44869\, -1.1455600 i & 3.81941\, -1.1443600 i \\
		\noalign{\smallskip}\hline
        0.40 & 0 & 2.70204\, -0.2675710 i & 3.11445\, -0.267572 i & 3.52750\, -0.2675380 i & 3.94102\, -0.2674920 i & 4.35467\, -0.2674610 i \\
        0.40 & 1 & 2.68907\, -0.8012110 i & 3.09450\, -0.804173 i & 3.50493\, -0.8049180 i & 3.92092\, -0.8042620 i & 4.33660\, -0.8038280 i \\
        0.40 & 2 & 2.76512\, -1.2786700 i & 3.09437\, -1.328090 i & 3.45897\, -1.3499700 i & 3.88047\, -1.3465800 i & 4.30133\, -1.3443700 i \\
       \noalign{\smallskip}\hline
        0.45 & 0 & 3.07962\, -0.3186010 i & 3.55034\, -0.318494 i & 4.02105\, -0.3184140 i & 4.49203\, -0.3183710 i & 4.96330\, -0.3183260 i \\
        0.45 & 1 & 3.04370\, -0.9602120 i & 3.52114\, -0.958682 i & 3.99411\, -0.9578740 i & 4.46649\, -0.9575940 i & 4.94048\, -0.9569200 i \\
        0.45 & 2 & 2.98794\, -1.6045500 i & 3.47342\, -1.604500 i & 3.94534\, -1.6031500 i & 4.41119\, -1.6061900 i & 4.89288\, -1.6020900 i \\
		\noalign{\smallskip}\hline
        0.50 & 0 & 3.55368\, -0.3858310 i & 4.09776\, -0.385525 i & 4.64146\, -0.3853430 i & 5.18474\, -0.3852710 i & 5.72839\, -0.3852120 i \\
        0.50 & 1 & 3.46324\, -1.1788800 i & 4.04730\, -1.164840 i & 4.60709\, -1.1597200 i & 5.15263\, -1.1590300 i & 5.70017\, -1.1580900 i \\
        0.50 & 2 & 2.97679\, -2.2509200 i & 3.88780\, -2.001720 i & 4.54783\, -1.9419300 i & 5.08584\, -1.9437800 i & 5.64601\, -1.9377600 i \\
		\noalign{\smallskip}\hline
	\end{tabular}
\end{table*}

\begin{table*}
    \centering
	\caption{Quasinormal frequencies of massless scalar perturbations computed using the Poschl-Teller fitting approach for $M = 1$ and $Q = 1/50$, with varying values of $\{\ell, n, l_b\}$.}
	\label{tab:1B}       
	\begin{tabular}{c|c|c|c|c|c|c}
		\hline\noalign{\smallskip}
		$\ell$  & $n$ & $l_b=6$ & $l_b=7$ & $l_b=8$ & $l_b=9$ & $l_b=10$  \\
		\noalign{\smallskip}\hline\noalign{\smallskip}
          \noalign{\smallskip}\hline
        0.00 & 0 & 1.25347\, -0.0964797 i & 1.44560\, -0.0964169 i & 1.63782\, -0.0963750 i & 1.83008\, -0.0963455 i & 2.02239\, -0.0963241 i \\
        0.00 & 1 & 1.25347\, -0.2894390 i & 1.44560\, -0.2892510 i & 1.63782\, -0.2891250 i & 1.83008\, -0.2890370 i & 2.02239\, -0.2889720 i \\
        0.00 & 2 & 1.25347\, -0.4823980 i & 1.44560\, -0.4820850 i & 1.63782\, -0.4818750 i & 1.83008\, -0.4817280 i & 2.02239\, -0.4816200 i \\
        \noalign{\smallskip}\hline    
        0.05 & 0 & 1.35408\, -0.1069180 i & 1.56154\, -0.1068450 i & 1.76909\, -0.1067960 i & 1.97671\, -0.1067610 i & 2.18437\, -0.1067360 i \\
        0.05 & 1 & 1.35408\, -0.3207530 i & 1.56154\, -0.3205340 i & 1.76909\, -0.3203870 i & 1.97671\, -0.3202840 i & 2.18437\, -0.3202090 i \\
        0.05 & 2 & 1.35408\, -0.5345880 i & 1.56154\, -0.5342230 i & 1.76909\, -0.5339780 i & 1.97671\, -0.5338070 i & 2.18437\, -0.5336820 i \\  
		\noalign{\smallskip}\hline
        0.10 & 0 & 1.46891\, -0.1191460 i & 1.69385\, -0.1190600 i & 1.91889\, -0.1190030 i & 2.14402\, -0.1189620 i & 2.36920\, -0.1189330 i \\
        0.10 & 1 & 1.46891\, -0.3574380 i & 1.69385\, -0.3571800 i & 1.91889\, -0.3570080 i & 2.14402\, -0.3568870 i & 2.36920\, -0.3567990 i \\
        0.10 & 2 & 1.46891\, -0.5957300 i & 1.69385\, -0.5953000 i & 1.91889\, -0.5950130 i & 2.14402\, -0.5948110 i & 2.36920\, -0.5946640 i \\
        \noalign{\smallskip}\hline
        0.15 & 0 & 1.60095\, -0.1335990 i & 1.84596\, -0.1334970 i & 2.09110\, -0.1334290 i & 2.33634\, -0.1333810 i & 2.58166\, -0.1333460 i \\
        0.15 & 1 & 1.60095\, -0.4007960 i & 1.84596\, -0.4004910 i & 2.09110\, -0.4002860 i & 2.33634\, -0.4001430 i & 2.58166\, -0.4000380 i \\
        0.15 & 2 & 1.60095\, -0.6679940 i & 1.84596\, -0.6674850 i & 2.09110\, -0.6671440 i & 2.33634\, -0.6669040 i & 2.58166\, -0.6667300 i \\
 	\noalign{\smallskip}\hline
        0.20 & 0 & 1.75403\, -0.1508500 i & 2.02228\, -0.1507280 i & 2.29071\, -0.1506460 i & 2.55925\, -0.1505890 i & 2.82788\, -0.1505470 i \\
        0.20 & 1 & 1.75403\, -0.4525510 i & 2.02228\, -0.4521850 i & 2.29071\, -0.4519390 i & 2.55925\, -0.4517670 i & 2.82788\, -0.4516420 i \\
        0.20 & 2 & 1.75403\, -0.7542510 i & 2.02228\, -0.7536410 i & 2.29071\, -0.7532320 i & 2.55925\, -0.7529450 i & 2.82788\, -0.7527360 i \\
        \noalign{\smallskip}\hline
        0.25 & 0 & 1.93316\, -0.1716730 i & 2.22859\, -0.1715250 i & 2.52423\, -0.1714250 i & 2.82002\, -0.1713560 i & 3.11591\, -0.1713050 i \\
        0.25 & 1 & 1.93316\, -0.5150180 i & 2.22859\, -0.5145740 i & 2.52423\, -0.5142760 i & 2.82002\, -0.5140680 i & 3.11591\, -0.5139160 i \\
        0.25 & 2 & 1.93316\, -0.8583630 i & 2.22859\, -0.8576230 i & 2.52423\, -0.8571270 i & 2.82002\, -0.8567800 i & 3.11591\, -0.8565260 i \\
	\noalign{\smallskip}\hline
        0.30 & 0 & 2.14502\, -0.1971240 i & 2.47254\, -0.1969420 i & 2.80032\, -0.1968200 i & 3.12829\, -0.1967350 i & 3.45639\, -0.1966720 i \\
        0.30 & 1 & 2.14502\, -0.5913720 i & 2.47254\, -0.5908260 i & 2.80032\, -0.5904610 i & 3.12829\, -0.5902040 i & 3.45639\, -0.5900170 i \\
        0.30 & 2 & 2.14502\, -0.9856190 i & 2.47254\, -0.9847100 i & 2.80032\, -0.9841020 i & 3.12829\, -0.9836740 i & 3.45639\, -0.9833620 i \\
		\noalign{\smallskip}\hline
        0.35 & 0 & 2.39862\, -0.2286850 i & 2.76450\, -0.2284590 i & 3.13071\, -0.2283070 i & 3.49715\, -0.2282000 i & 3.86377\, -0.2281220 i \\
        0.35 & 1 & 2.39862\, -0.6860560 i & 2.76450\, -0.6853760 i & 3.13071\, -0.6849200 i & 3.49715\, -0.6846000 i & 3.86377\, -0.6843670 i \\
        0.35 & 2 & 2.39862\, -1.1434300 i & 2.76450\, -1.1422900 i & 3.13071\, -1.1415300 i & 3.49715\, -1.1410000 i & 3.86377\, -1.1406100 i \\
		\noalign{\smallskip}\hline
        0.40 & 0 & 2.70648\, -0.2684820 i & 3.11883\, -0.2681940 i & 3.53162\, -0.2680010 i & 3.94470\, -0.2678660 i & 4.35800\, -0.2677670 i \\
        0.40 & 1 & 2.70648\, -0.8054450 i & 3.11883\, -0.8045820 i & 3.53162\, -0.8040040 i & 3.94470\, -0.8035970 i & 4.35800\, -0.8033010 i \\
        0.40 & 2 & 2.70648\, -1.3424100 i & 3.11883\, -1.3409700 i & 3.53162\, -1.3400100 i & 3.94470\, -1.3393300 i & 4.35800\, -1.3388300 i \\
       \noalign{\smallskip}\hline
        0.45 & 0 & 3.08636\, -0.3196490 i & 3.55595\, -0.3192760 i & 4.02610\, -0.3190260 i & 4.49664\, -0.3188510 i & 4.96745\, -0.3187230 i \\
        0.45 & 1 & 3.08636\, -0.9589460 i & 3.55595\, -0.9578290 i & 4.02610\, -0.9570790 i & 4.49664\, -0.9565520 i & 4.96745\, -0.9561680 i \\
        0.45 & 2 & 3.08636\, -1.5982400 i & 3.55595\, -1.5963800 i & 4.02610\, -1.5951300 i & 4.49664\, -1.5942500 i & 4.96745\, -1.5936100 i \\
		\noalign{\smallskip}\hline
        0.50 & 0 & 3.56432\, -0.3869710 i & 4.10573\, -0.3864770 i & 4.64790\, -0.3861450 i & 5.19057\, -0.3859110 i & 5.73362\, -0.3857410 i \\
        0.50 & 1 & 3.56432\, -1.1609100 i & 4.10573\, -1.1594300 i & 4.64790\, -1.1584300 i & 5.19057\, -1.1577300 i & 5.73362\, -1.1572200 i \\
        0.50 & 2 & 3.56432\, -1.9348600 i & 4.10573\, -1.9323800 i & 4.64790\, -1.9307200 i & 5.19057\, -1.9295600 i & 5.73362\, -1.9287000 i \\
		\noalign{\smallskip}\hline
	\end{tabular}
\end{table*}

\begin{table*}
    \centering
	\caption{Quasinormal frequencies of massless Dirac perturbations computed using the 3rd-order WKB approximation for $M = 1$ and $Q = 1/50$, with varying values of $\{\ell, n,\xi\}$.}
	\label{tab:2}       
	\begin{tabular}{c|c|c|c|c|c|c}
		\hline\noalign{\smallskip}
		$\ell$  & $n$ & $\xi=6$ & $\xi=7$ & $\xi=8$ & $\xi=9$ & $\xi=10$  \\
		\noalign{\smallskip}\hline\noalign{\smallskip}
          \noalign{\smallskip}\hline
        0.00 & 0 & 1.15310\, -0.0962484 i & 1.34582\, -0.0962426 i & 1.53846\, -0.0962388 i & 1.73107\, -0.0962363 i & 1.92364\, -0.0962345 i \\
        0.00 & 1 & 1.14424\, -0.2897200 i & 1.33821\, -0.2894420 i & 1.53180\, -0.2892630 i & 1.72514\, -0.2891400 i & 1.91831\, -0.2890530 i \\
        0.00 & 2 & 1.12749\, -0.4857700 i & 1.32363\, -0.4845950 i & 1.51891\, -0.4838130 i & 1.71360\, -0.4832670 i & 1.90786\, -0.4828720 i \\
        \noalign{\smallskip}\hline    
        0.05 & 0 & 1.24524\, -0.1066480 i & 1.45338\, -0.1066410 i & 1.66145\, -0.1066370 i & 1.86947\, -0.1066340 i & 2.07745\, -0.1066320 i \\
        0.05 & 1 & 1.23516\, -0.3210820 i & 1.44474\, -0.3207580 i & 1.65388\, -0.3205480 i & 1.86274\, -0.3204050 i & 2.07139\, -0.3203030 i \\
        0.05 & 2 & 1.21618\, -0.5385050 i & 1.42820\, -0.5371420 i & 1.63925\, -0.5362330 i & 1.84963\, -0.5355990 i & 2.05953\, -0.5351390 i \\ 
		\noalign{\smallskip}\hline
        0.10 & 0 & 1.35034\, -0.1188290 i & 1.57609\, -0.1188210 i & 1.80175\, -0.1188160 i & 2.02735\, -0.1188130 i & 2.25292\, -0.1188100 i \\
        0.10 & 1 & 1.33881\, -0.3578260 i & 1.56619\, -0.3574450 i & 1.79308\, -0.3571980 i & 2.01965\, -0.3570300 i & 2.24598\, -0.3569100 i \\
        0.10 & 2 & 1.31715\, -0.6003150 i & 1.54731\, -0.5987210 i & 1.77637\, -0.5976580 i & 2.00467\, -0.5969140 i & 2.23243\, -0.5963750 i \\
		\noalign{\smallskip}\hline
        0.15 & 0 & 1.47110\, -0.1332230 i & 1.71709\, -0.1332140 i & 1.96297\, -0.1332070 i & 2.20879\, -0.1332030 i & 2.45456\, -0.1332000 i \\
        0.15 & 1 & 1.45780\, -0.4012590 i & 1.70567\, -0.4008050 i & 1.95297\, -0.4005130 i & 2.19990\, -0.4003130 i & 2.44656\, -0.4001700 i \\
        0.15 & 2 & 1.43291\, -0.6734100 i & 1.68395\, -0.6715300 i & 1.93374\, -0.6702730 i & 2.18266\, -0.6693940 i & 2.43094\, -0.6687570 i \\
		\noalign{\smallskip}\hline
        0.20 & 0 & 1.61101\, -0.1504000 i & 1.88044\, -0.1503890 i & 2.14976\, -0.1503810 i & 2.41900\, -0.1503760 i & 2.68819\, -0.1503730 i \\
        0.20 & 1 & 1.59553\, -0.4531090 i & 1.86716\, -0.4525630 i & 2.13813\, -0.4522120 i & 2.40866\, -0.4519720 i & 2.67888\, -0.4518010 i \\
        0.20 & 2 & 1.56669\, -0.7607110 i & 1.84197\, -0.7584720 i & 2.11580\, -0.7569730 i & 2.38863\, -0.7559240 i & 2.66074\, -0.7551620 i \\
        \noalign{\smallskip}\hline
        0.25 & 0 & 1.77459\, -0.1711270 i & 2.07146\, -0.1711130 i & 2.36818\, -0.1711040 i & 2.66482\, -0.1710980 i & 2.96140\, -0.1710930 i \\
        0.25 & 1 & 1.75641\, -0.5156990 i & 2.05586\, -0.5150360 i & 2.35452\, -0.5146080 i & 2.65267\, -0.5143170 i & 2.95045\, -0.5141090 i \\
        0.25 & 2 & 1.72267\, -0.8661540 i & 2.02635\, -0.8634570 i & 2.32836\, -0.8616500 i & 2.62919\, -0.8603830 i & 2.92918\, -0.8594620 i \\
	\noalign{\smallskip}\hline
        0.30 & 0 & 1.96786\, -0.1964540 i & 2.29716\, -0.1964360 i & 2.62628\, -0.1964250 i & 2.95530\, -0.1964170 i & 3.28425\, -0.1964120 i \\
        0.30 & 1 & 1.94628\, -0.5922140 i & 2.27863\, -0.5913970 i & 2.61005\, -0.5908710 i & 2.94087\, -0.5905110 i & 3.27125\, -0.5902550 i \\
        0.30 & 2 & 1.90639\, -0.9951350 i & 2.24371\, -0.9918470 i & 2.57907\, -0.9896400 i & 2.91304\, -0.9880900 i & 3.24602\, -0.9869620 i \\
		\noalign{\smallskip}\hline
        0.35 & 0 & 2.19897\, -0.2278500 i & 2.56706\, -0.2278270 i & 2.93495\, -0.2278130 i & 3.30270\, -0.2278040 i & 3.67036\, -0.2277970 i \\
        0.35 & 1 & 2.17300\, -0.6871150 i & 2.54477\, -0.6860920 i & 2.91542\, -0.6854340 i & 3.28533\, -0.6849850 i & 3.65472\, -0.6846650 i \\
        0.35 & 2 & 2.12529\, -1.1552200 i & 2.50294\, -1.1511500 i & 2.87826\, -1.1484200 i & 3.25194\, -1.1464900 i & 3.62443\, -1.1450900 i \\
		\noalign{\smallskip}\hline
        0.40 & 0 & 2.47916\, -0.2674210 i & 2.89432\, -0.2673920 i & 3.30922\, -0.267374 i & 3.723960\, -0.2673620 i & 4.13858\, -0.2673540 i \\
        0.40 & 1 & 2.44746\, -0.8068030 i & 2.86710\, -0.8054990 i & 3.28538\, -0.804660 i & 3.702740\, -0.8040880 i & 4.11948\, -0.8036810 i \\
        0.40 & 2 & 2.38957\, -1.3572600 i & 2.81628\, -1.3521500 i & 3.24018\, -1.348700 i & 3.662100\, -1.3462800 i & 4.08259\, -1.3445100 i \\
       \noalign{\smallskip}\hline
        0.45 & 0 & 2.82440\, -0.3182750 i & 3.29759\, -0.3182370 i & 3.77045\, -0.318214 i & 4.243110\, -0.3181980 i & 4.71563\, -0.3181870 i \\
        0.45 & 1 & 2.78501\, -0.9607260 i & 3.26377\, -0.9590280 i & 3.74083\, -0.957936 i & 4.216750\, -0.9571930 i & 4.69189\, -0.9566640 i \\
        0.45 & 2 & 2.71364\, -1.6173200 i & 3.20099\, -1.6107700 i & 3.68493\, -1.606350 i & 4.166440\, -1.6032300 i & 4.64620\, -1.6009400 i \\
		\noalign{\smallskip}\hline
        0.50 & 0 & 3.25798\, -0.3851470 i & 3.80412\, -0.3850970 i & 4.34984\, -0.385065 i & 4.895300\, -0.3850440 i & 5.44058\, -0.3850290 i \\
        0.50 & 1 & 3.20804\, -1.1633100 i & 3.76123\, -1.1610400 i & 4.31227\, -1.159580 i & 4.861870\, -1.1585900 i & 5.41047\, -1.1578900 i \\
        0.50 & 2 & 3.11832\, -1.9599400 i & 3.68216\, -1.9513500 i & 4.24175\, -1.945530 i & 4.798330\, -1.9414200 i & 5.35271\, -1.9384100 i \\
		\noalign{\smallskip}\hline
	\end{tabular}
\end{table*}

\begin{table*}
    \centering
	\caption{Quasinormal frequencies of massless Dirac perturbations computed using the Poschl-Teller fitting approach for $M = 1$ and $Q = 1/50$, with varying values of $\{\ell, n,\xi\}$.}
	\label{tab:2B}       
	\begin{tabular}{c|c|c|c|c|c|c}
		\hline\noalign{\smallskip}
		$\ell$  & $n$ & $\xi=6$ & $\xi=7$ & $\xi=8$ & $\xi=9$ & $\xi=10$  \\
		\noalign{\smallskip}\hline\noalign{\smallskip}
          \noalign{\smallskip}\hline
        0.00 & 0 & 1.15499\, -0.0965009 i & 1.34740\, -0.0964234 i & 1.53983\, -0.0963745 i & 1.73227\, -0.0963417 i & 1.92471\, -0.0963187 i \\
        0.00 & 1 & 1.15499\, -0.2895030 i & 1.34740\, -0.2892700 i & 1.53983\, -0.2891230 i & 1.73227\, -0.2890250 i & 1.92471\, -0.2889560 i \\
        0.00 & 2 & 1.15499\, -0.4825050 i & 1.34740\, -0.4821170 i & 1.53983\, -0.4818720 i & 1.73227\, -0.4817090 i & 1.92471\, -0.4815940 i \\
        \noalign{\smallskip}\hline    
        0.05 & 0 & 1.24739\, -0.1069440 i & 1.45519\, -0.1068530 i & 1.66301\, -0.1067960 i & 1.87084\, -0.1067570 i & 2.07868\, -0.1067300 i \\
        0.05 & 1 & 1.24739\, -0.3208320 i & 1.45519\, -0.3205590 i & 1.66301\, -0.3203870 i & 1.87084\, -0.3202720 i & 2.07868\, -0.3201910 i \\
        0.05 & 2 & 1.24739\, -0.5347200 i & 1.45519\, -0.5342650 i & 1.66301\, -0.5339790 i & 1.87084\, -0.5337870 i & 2.07868\, -0.5336520 i \\
		\noalign{\smallskip}\hline
        0.10 & 0 & 1.35281\, -0.1191790 i & 1.57816\, -0.1190710 i & 1.80354\, -0.1190040 i & 2.02893\, -0.1189580 i & 2.25432\, -0.1189260 i \\
        0.10 & 1 & 1.35281\, -0.3575360 i & 1.57816\, -0.3572140 i & 1.80354\, -0.3570110 i & 2.02893\, -0.3568750 i & 2.25432\, -0.3567790 i \\
        0.10 & 2 & 1.35281\, -0.5958940 i & 1.57816\, -0.5953570 i & 1.80354\, -0.5950180 i & 2.02893\, -0.5947910 i & 2.25432\, -0.5946320 i \\
		\noalign{\smallskip}\hline
        0.15 & 0 & 1.47396\, -0.1336400 i & 1.71949\, -0.1335120 i & 1.96504\, -0.1334310 i & 2.21061\, -0.1333770 i & 2.45618\, -0.1333390 i \\
        0.15 & 1 & 1.47396\, -0.4009210 i & 1.71949\, -0.4005360 i & 1.96504\, -0.4002930 i & 2.21061\, -0.4001300 i & 2.45618\, -0.4000170 i \\
        0.15 & 2 & 1.47396\, -0.6682010 i & 1.71949\, -0.6675590 i & 1.96504\, -0.6671550 i & 2.21061\, -0.6668840 i & 2.45618\, -0.6666940 i \\
		\noalign{\smallskip}\hline
        0.20 & 0 & 1.61435\, -0.1509030 i & 1.88325\, -0.1507480 i & 2.15218\, -0.1506510 i & 2.42112\, -0.1505850 i & 2.69008\, -0.1505390 i \\
        0.20 & 1 & 1.61435\, -0.4527090 i & 1.88325\, -0.4522450 i & 2.15218\, -0.4519520 i & 2.42112\, -0.4517560 i & 2.69008\, -0.4516180 i \\
        0.20 & 2 & 1.61435\, -0.7545150 i & 1.88325\, -0.7537410 i & 2.15218\, -0.7532530 i & 2.42112\, -0.7529260 i & 2.69008\, -0.7526970 i \\
        \noalign{\smallskip}\hline
        0.25 & 0 & 1.77853\, -0.1717410 i & 2.07477\, -0.1715520 i & 2.37103\, -0.1714320 i & 2.66732\, -0.1713530 i & 2.96362\, -0.1712970 i \\
        0.25 & 1 & 1.77853\, -0.5152220 i & 2.07477\, -0.5146550 i & 2.37103\, -0.5142970 i & 2.66732\, -0.5140580 i & 2.96362\, -0.5138900 i \\
        0.25 & 2 & 1.77853\, -0.8587040 i & 2.07477\, -0.8577580 i & 2.37103\, -0.8571620 i & 2.66732\, -0.8567630 i & 2.96362\, -0.8564840 i \\
	\noalign{\smallskip}\hline
        0.30 & 0 & 1.97258\, -0.1972130 i & 2.30111\, -0.1969790 i & 2.62969\, -0.1968310 i & 2.95828\, -0.1967330 i & 3.28690\, -0.1966630 i \\
        0.30 & 1 & 1.97258\, -0.5916400 i & 2.30111\, -0.5909370 i & 2.62969\, -0.5904940 i & 2.95828\, -0.5901980 i & 3.28690\, -0.5899900 i \\
        0.30 & 2 & 1.97258\, -0.9860670 i & 2.30111\, -0.9848950 i & 2.62969\, -0.9841560 i & 2.95828\, -0.9836630 i & 3.28690\, -0.9833170 i \\
		\noalign{\smallskip}\hline
        0.35 & 0 & 2.20467\, -0.2288050 i & 2.57184\, -0.2285100 i & 2.93906\, -0.2283240 i & 3.30630\, -0.2282000 i & 3.67357\, -0.2281130 i \\
        0.35 & 1 & 2.20467\, -0.6864140 i & 2.57184\, -0.6855290 i & 2.93906\, -0.6849720 i & 3.30630\, -0.6845990 i & 3.67357\, -0.6843380 i \\
        0.35 & 2 & 2.20467\, -1.1440200 i & 2.57184\, -1.1425500 i & 2.93906\, -1.1416200 i & 3.30630\, -1.1410000 i & 3.67357\, -1.1405600 i \\
		\noalign{\smallskip}\hline
        0.40 & 0 & 2.48617\, -0.2686440 i & 2.90019\, -0.2682660 i & 3.31427\, -0.2680280 i & 3.72838\, -0.2678690 i & 4.14252\, -0.2677580 i \\
        0.40 & 1 & 2.48617\, -0.8059320 i & 2.90019\, -0.8047990 i & 3.31427\, -0.8040840 i & 3.72838\, -0.8036070 i & 4.14252\, -0.8032730 i \\
        0.40 & 2 & 2.48617\, -1.3432200 i & 2.90019\, -1.3413300 i & 3.31427\, -1.3401400 i & 3.72838\, -1.3393400 i & 4.14252\, -1.3387900 i \\
       \noalign{\smallskip}\hline
        0.45 & 0 & 2.83317\, -0.3198750 i & 3.30493\, -0.3193800 i & 3.77676\, -0.3190690 i & 4.24864\, -0.3188600 i & 4.72054\, -0.3187140 i \\
        0.45 & 1 & 2.83317\, -0.9596260 i & 3.30493\, -0.9581410 i & 3.77676\, -0.9572060 i & 4.24864\, -0.9565800 i & 4.72054\, -0.9561430 i \\
        0.45 & 2 & 2.83317\, -1.5993800 i & 3.30493\, -1.5969000 i & 3.77676\, -1.5953400 i & 4.24864\, -1.5943000 i & 4.72054\, -1.5935700 i \\
		\noalign{\smallskip}\hline
        0.50 & 0 & 3.26920\, -0.3872950 i & 3.81351\, -0.3866310 i & 4.35790\, -0.3862120 i & 4.90235\, -0.3859320 i & 5.44684\, -0.3857360 i \\
        0.50 & 1 & 3.26920\, -1.1618900 i & 3.81351\, -1.1598900 i & 4.35790\, -1.1586400 i & 4.90235\, -1.1578000 i & 5.44684\, -1.1572100 i \\
        0.50 & 2 & 3.26920\, -1.9364800 i & 3.81351\, -1.9331500 i & 4.35790\, -1.9310600 i & 4.90235\, -1.9296600 i & 5.44684\, -1.9286800 i \\
		\noalign{\smallskip}\hline
	\end{tabular}
\end{table*}

\section{The energy deposition rate by the neutrino annihilation process}

In this section, we consider the energy deposition rate in the spacetime governed by a black hole, accounting for mass deformation effects. The energy deposition per unit time and per unit volume due to neutrino pair annihilation can be expressed as \cite{Salmonson:1999es}:
\begin{align}
\frac{\mathrm{d}E(\mathbf{r})}{\mathrm{d}t\mathrm{d}V}=2KG_{F}^{2}F(r)\iint
n(\varepsilon_{\nu})n(\varepsilon_{\overline{\nu}})
(\varepsilon_{\nu}+\varepsilon_{\overline{\nu}})
\varepsilon_{\nu}^{3}\varepsilon_{\overline{\nu}}^{3}
\mathrm{d}\varepsilon_{\nu}\mathrm{d}\varepsilon_{\overline{\nu}}.
\end{align}
Here, the constant $K$ encapsulates the weak interaction strength and is explicitly dependent on the neutrino species through the Weinberg angle $\sin^{2}\theta_{W}=0.23$. Its forms are specifically given by:
\begin{align}
K(\nu_{\mu},\overline{\nu}_{\mu})=K(\nu_{\tau},\overline{\nu}_{\tau})&=
\frac{1}{6\pi}\left(1-4\sin^{2}\theta_{W}+8\sin^{4}\theta_{W}\right),\\[8pt]
K(\nu_{e},\overline{\nu}_{e})&=
\frac{1}{6\pi}\left(1+4\sin^{2}\theta_{W}+8\sin^{4}\theta_{W}\right).
\end{align}
The numerical value of the Fermi constant is $G_{F}=5.29\times 10^{-44}\,\mathrm{cm}^{2}\mathrm{MeV}^{-2}$, characterizing the weak interaction.

The angular dependence is encapsulated in the factor $F(r)$:
\begin{align}
F(r)=\frac{2\pi^{2}}{3}(1-x)^{4}(x^{2}+4x+5),
\end{align}
where the variable $x=\sin\theta_{r}$ denotes the angle between the neutrino trajectory and the tangent vector of a circular orbit at radius $r$. This angular factor emphasizes the geometric structure and orientation dependence of the neutrino annihilation process.

The neutrino and antineutrino distributions, $n(\varepsilon_{\nu})$ and $n(\varepsilon_{\overline{\nu}})$, follow the Fermi-Dirac distribution at local temperature $T(r)$:
\begin{align}
n(\varepsilon_{\nu})=\frac{2}{h^{3}}\frac{1}{\exp\left(\frac{\varepsilon_{\nu}}{kT}\right)+1}.
\end{align}
Integration over the energy space yields the energy deposition per unit volume:
\begin{align}
\frac{\mathrm{d}E}{\mathrm{d}t\mathrm{d}V}=\frac{21\zeta(5)\pi^{4}}{h^{6}}KG_{F}^{2}F(r)(kT)^{9},
\end{align}
where the temperature distribution must consider gravitational redshift effects. Local temperature and luminosity relate through:
\begin{align}
T(r)\sqrt{g_{00}(r)}&=T(R)\sqrt{g_{00}(R)},\\[8pt]
L_{\infty}&=g_{00}(R)L(R).
\end{align}

Consequently, the energy deposition rate can be expressed as:
\begin{align}
\frac{\mathrm{d}E(\mathbf{r})}{\mathrm{d}t\mathrm{d}V}=\frac{21\zeta(5)\pi^{4}}{h^{6}}KG_{F}^{2}k^{9}\left(\frac{7}{4}\pi ac\right)^{-\frac{9}{4}}L_{\infty}^{\frac{9}{4}}F(r)\frac{\left[g_{00}(R)\right]^{\frac{9}{4}}}{\left[g_{00}(r)\right]^{\frac{9}{4}}}R_{0}^{-\frac{9}{2}}.
\end{align}

To proceed, we require explicit expressions for the angular variable $x$, connected to the spacetime metric as follows \cite{Lambiase:2020iul,Shi:2023kid}:
\begin{align}
x^{2}=1-\frac{R^{2}}{r^{2}}\frac{f(r)}{f(R)}, \quad \text{with}\quad f(r)=g_{00}(r).
\end{align}

The total energy conversion rate from neutrino annihilation, integrated radially, is:
\begin{align}
\dot{Q}=\frac{84\zeta(5)\pi^{5}}{h^{6}}KG_{F}^{2}k^{9}\left(\frac{7}{4}\pi ac\right)^{-\frac{9}{4}}L_{\infty}^{\frac{9}{4}}\left[g_{00}(R)\right]^{\frac{9}{4}}R^{-\frac{9}{2}}\int_{R_{0}}^{\infty}\frac{r^{2}\sqrt{-g_{11}(r)}F(r)}{g_{00}(r)}dr.
\end{align}

The ratio of general relativistic to Newtonian deposition rates is crucial for comparison and interpretation of gravitational effects:
\begin{align}
\frac{\dot{Q}}{\dot{Q}_{\mathrm{Newt}}}=3\left[g_{00}(R)\right]^{\frac{9}{4}}\int_{1}^{\infty}(x-1)^{4}(x^{2}+4x+5)\frac{y^{2}\sqrt{-g_{11}(Ry)}}{g_{00}(Ry)^{\frac{9}{2}}}dy.
\end{align}

\begin{figure}[t]
\includegraphics[width=10cm]{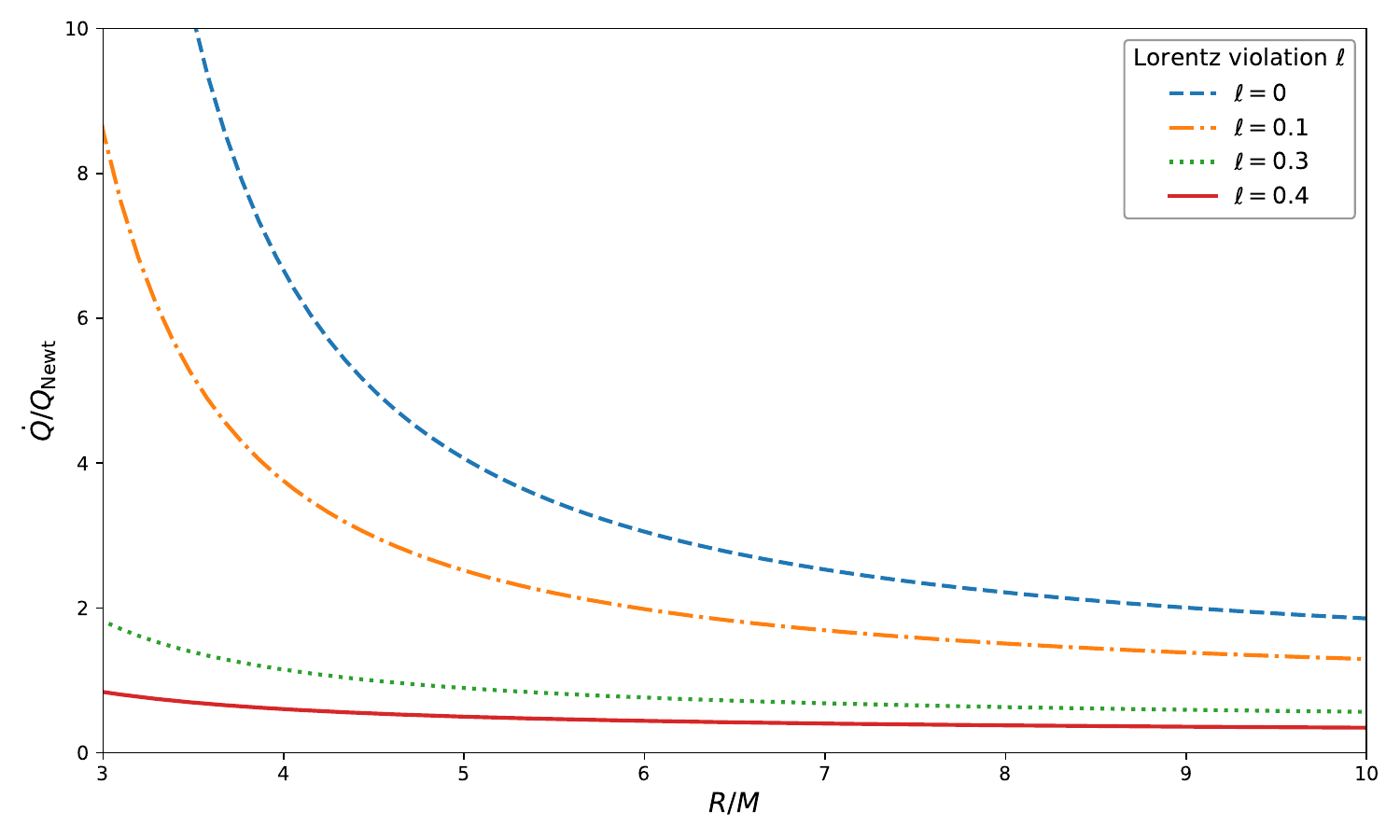}
\label{neutrino}
\caption{The solid, dotted and dashed curves of the ratio $\dfrac{\dot{Q}}{\dot{Q_{Newt}}}$ as functions of the ratio $\dfrac{R}{M}$ for the parameter $\ell=0, 0.1, 0.3, 0.4$ respectively.}
\end{figure}

Figure \ref{neutrino} illustrates how the energy deposition rate ratio $\dot{Q}/\dot{Q}_{\mathrm{Newt}}$ varies with the dimensionless parameter $R/M$ for different values of the Lorentz violation parameter $\ell=0, 0.1, 0.3, 0.4$. It highlights how deviations from classical gravity predictions increase as Lorentz violation becomes more pronounced. This graphical representation provides insight into how spacetime deformation affects neutrino annihilation efficiency around gravitational objects.


\section{Conclusion}
In this paper, we also derived the deflection angle and black hole shadow characteristics for an electrically charged black hole immersed in a Lorentz-violating background sourced by a Kalb-Ramond (KR) antisymmetric tensor field. The analytical treatment of the deflection angle employs the Gauss-Bonnet theorem in a non-asymptotically flat spacetime, allowing for a rigorous extraction of light bending effects under the influence of a nontrivial KR-induced background geometry. The resulting expressions reveal that the Lorentz-violating parameter $ \ell $, intimately tied to the field strength of the KR background and its coupling to the electromagnetic sector, introduces corrections that persist even in the null-charge limit—thereby encoding LIV effects into purely geometric deformations of spacetime. Notably, the deflection angle exhibits a pronounced dependence on particle velocity, especially in the non-relativistic regime where a $ 1/v^2 $ divergence emerges, hinting at the utility of low-energy lensing as a potential probe of spontaneous Lorentz symmetry breaking. The shadow analysis demonstrates a nontrivial reshaping of the photon sphere and critical impact parameter, culminating in a shadow radius expression that not only deviates from the classical Reissner–Nordström profile but also displays an electrodynamic-geometric interplay due to the non-minimal coupling. Particularly significant is the emergence of a term in the shadow radius that can invert the expected charge dependence for small positive $ \ell $, an effect absent in standard Einstein-Maxwell theory. The constraints on $ \ell $ obtained from Event Horizon Telescope data further corroborate the theoretical consistency and observational relevance of the model, placing the parameter within physically plausible bounds.

We have also studied both scalar and Dirac massless quasinormal modes in an electrically charged black hole with a Kalb-Ramond background field in four dimensions. After introducing the importance of QNMs, the key ideas and the formalism used to obtain them, we show in figures the effective potential for each type of perturbation considered. We have noticed that the potential has a convenient form after varying all the parameters, which is the reason why we have used the high-order WBK approximation to obtain the corresponding QNMs. 
As a complementary check, we have implemented the well-known Poschl-Teller fitting approach to approximate the effective potential by the Poschl-Teller and subsequently solve the Shrodinger-like differential equation to obtain the quasinormal frequencies. As has been pointed out, this method is only useful in certain cases, in particular for low values of $n$ and high values of $l_b$ (or equivalently $\xi$).
Since the imaginary part of the QN frequencies is negative, this fact shows that the black hole is stable against such perturbations. To illustrate this, we have added some figures to visualise the results. 
At this point, it is essential to point out that there are some interesting works where other authors have computed QNMs of this and similar problems, using different methods and with a cosmological constant with electric charge (see \cite{Hu:2025isj,Zahid:2024hyy,AraujoFilho:2024rcr,Guo:2023nkd}). In our paper,, we have focused on higher order QNMs using the WKB approximation and we have found that, at 10th order, the computation is stable and the results are consistent. In addition, we have for the first time calculated Dirac QN frequencies for the massless case using the 3rd-order WKB approximation. Higher orders are possible, but the routine is time-consuming and the 3rd order is sufficient.

Lastly,  we investigated the neutrino pair annihilation \(\nu\bar{\nu}\rightarrow e^-e^+\) around charged black holes within the framework of Kalb-Ramond gravity. We derived detailed analytical expressions for the energy deposition rate per unit time and volume, emphasizing the dependence on neutrino temperature distributions, angular integrations, and spacetime metric parameters. The obtained equations explicitly show how the metric functions significantly influence the annihilation efficiency, thus affecting potential astrophysical phenomena such as GRBs. The ratio \(\frac{\dot{Q}}{\dot{Q}_{Newt}}\), representing the enhancement of energy deposition compared to the Newtonian case, was thoroughly analyzed numerically. Our results, illustrated in Fig. 1, clearly demonstrate how the parameter \(\ell\), associated with Kalb-Ramond gravity corrections, strongly affects the neutrino annihilation efficiency. An increase in \(\ell\) corresponds to a pronounced amplification of the annihilation rate, especially noticeable at smaller radial distances from the black hole, indicating that gravitational effects beyond general relativity may play a crucial role in powering GRBs.

Moreover, we explored the radial dependence of energy deposition through \(\frac{\mathrm{d}\dot{Q}}{\mathrm{d}r}\), highlighting regions near the black hole surface as particularly significant in terms of energy production. Our analysis suggests that compact astrophysical objects described by Kalb-Ramond gravity provide sufficiently intense neutrino annihilation conditions, potentially explaining observational data from gamma-ray bursts. This study underscores the importance of exploring gravitational modifications beyond standard general relativity to better understand extreme astrophysical phenomena. Future observational studies of GRBs may offer opportunities to test and constrain theories like Kalb-Ramond gravity, ultimately enhancing our understanding of fundamental physics in strong gravitational fields.

\acknowledgments

A.~R. acknowledge financial support from Conselleria d'Educació, Cultura, Universitats i Ocupació de la Generalitat Valenciana thorugh PROMETEO PROJECT CIPROM/2022/13.
A.~R. is very grateful for the hospitality of the Centre for Theoretical Physics and Astrophysics, Institute of Physics, Silesian University in Opava, the University of Valencia, and the Complutense University of Madrid, in Spain.
R. P. and A. O. would like to acknowledge networking support of the COST Action CA18108 - Quantum gravity phenomenology in the multi-messenger approach (QG-MM), COST Action CA21106 - COSMIC WISPers in the Dark Universe: Theory, astrophysics and experiments (CosmicWISPers), the COST Action CA22113 - Fundamental challenges in theoretical physics (THEORY-CHALLENGES), the COST Action CA21136 - Addressing observational tensions in cosmology with systematics and fundamental physics (CosmoVerse), the COST Action CA23130 - Bridging high and low energies in search of quantum gravity (BridgeQG), and the COST Action CA23115 - Relativistic Quantum Information (RQI) funded by COST (European Cooperation in Science and
Technology). A. O also thanks to EMU, TUBITAK, ULAKBIM (Turkiye) and SCOAP3 (Switzerland) for their support.  
\bibliography{references.bib}
\end{document}